\PassOptionsToPackage{unicode}{hyperref}
\PassOptionsToPackage{hyphens}{url}
\documentclass[
]{article}
\usepackage{lmodern}
\usepackage{amssymb,amsmath}
\usepackage{ifxetex,ifluatex}
\ifnum 0\ifxetex 1\fi\ifluatex 1\fi=0 % if pdftex
  \usepackage[T1]{fontenc}
  \usepackage[utf8]{inputenc}
  \usepackage{textcomp} % provide euro and other symbols
\else % if luatex or xetex
  \usepackage{unicode-math}
  \defaultfontfeatures{Scale=MatchLowercase}
  \defaultfontfeatures[\rmfamily]{Ligatures=TeX,Scale=1}
\fi
\IfFileExists{upquote.sty}{\usepackage{upquote}}{}
\IfFileExists{microtype.sty}{% use microtype if available
  \usepackage[]{microtype}
  \UseMicrotypeSet[protrusion]{basicmath} % disable protrusion for tt fonts
}{}
\makeatletter
\@ifundefined{KOMAClassName}{% if non-KOMA class
  \IfFileExists{parskip.sty}{%
    \usepackage{parskip}
  }{% else
    \setlength{\parindent}{0pt}
    \setlength{\parskip}{6pt plus 2pt minus 1pt}}
}{% if KOMA class
  \KOMAoptions{parskip=half}}
\makeatother
\usepackage{xcolor}
\IfFileExists{xurl.sty}{\usepackage{xurl}}{} % add URL line breaks if available
\IfFileExists{bookmark.sty}{\usepackage{bookmark}}{\usepackage{hyperref}}
\hypersetup{
  pdftitle={Tree species, crown cover, and age as determinants of the vertical distribution of airborne LiDAR returns},
  pdfauthor={Etienne Racine1,*, Nicholas C. Coops2, Jean Bégin1, Mari Myllymäki3},
  hidelinks,
  pdfcreator={LaTeX via pandoc}}
\usepackage[margin=1in]{geometry}
\usepackage{longtable,booktabs}
\usepackage{etoolbox}
\makeatletter
\patchcmd\longtable{\par}{\if@noskipsec\mbox{}\fi\par}{}{}
\makeatother
\IfFileExists{footnotehyper.sty}{\usepackage{footnotehyper}}{\usepackage{footnote}}
\makesavenoteenv{longtable}
\usepackage{graphicx,grffile}
\makeatletter
\def\maxwidth{\ifdim\Gin@nat@width>\linewidth\linewidth\else\Gin@nat@width\fi}
\def\maxheight{\ifdim\Gin@nat@height>\textheight\textheight\else\Gin@nat@height\fi}
\makeatother
\setkeys{Gin}{width=\maxwidth,height=\maxheight,keepaspectratio}
\makeatletter
\def\fps@figure{htbp}
\makeatother
\usepackage[utf8]{inputenc}
\usepackage{amssymb}
\usepackage{pifont}
\usepackage{newunicodechar}
\usepackage{gensymb}
\usepackage{longtable,tabu} % longtabu
\usepackage[export]{adjustbox} % valign
\DeclareUnicodeCharacter{211D}{\mathbb{R}}
\DeclareUnicodeCharacter{2009}{\,}
\DeclareUnicodeCharacter{221E}{\ensuremath{\infty}}
\DeclareUnicodeCharacter{2300}{\ensuremath{\oslash}}
\DeclareUnicodeCharacter{0394}{\ensuremath{\Delta}}
\DeclareUnicodeCharacter{2212}{\textminus}
\DeclareUnicodeCharacter{2248}{\ensuremath{\approx}}
\DeclareUnicodeCharacter{2265}{\ensuremath{\geq}}
\DeclareUnicodeCharacter{2264}{\ensuremath{\leq}}
\DeclareUnicodeCharacter{2800}{\,}
\DeclareUnicodeCharacter{200B}{\,}
\newunicodechar{²}{\ensuremath{{}^2}}
\newunicodechar{✓}{\ding{51}}

\title{Tree species, crown cover, and age as determinants of the vertical distribution of airborne LiDAR returns}
\author{Etienne Racine\textsuperscript{1,*}, Nicholas C. Coops\textsuperscript{2}, Jean Bégin\textsuperscript{1}, Mari Myllymäki\textsuperscript{3}}
\date{\textsuperscript{1}:Département des sciences du bois et de la forêt, 2405 rue de la Terrasse, Université Laval, Québec, QC G1V 0A6, Canada; \textsuperscript{2}:Department of Forest Resources Management, 2424 Main Mall, University of British Columbia, Vancouver, BC V6T 1Z4, Canada; \textsuperscript{3}:Natural Resources Institute Finland (Luke), Latokartanonkaari 9, FI-00790 Helsinki, Finland}

\begin{document}
\maketitle


\hypertarget{key-message}{%
\section*{Key Message}\label{key-message}}
\addcontentsline{toc}{section}{Key Message}

We assessed even-aged stand vertical distributions of LiDAR returns and found that tree species, age, and crown cover each have a distinct pattern that together explain up to 47\% of the variation.

\hypertarget{abstract}{%
\section*{Abstract}\label{abstract}}
\addcontentsline{toc}{section}{Abstract}

Light detection and ranging (LiDAR) provides information on the vertical structure of forest stands enabling detailed and extensive ecosystem study.
The vertical structure is often summarized by scalar features and data-reduction techniques that limit the interpretation of results.
Instead, we quantified the influence of three variables, species, crown cover, and age, on the vertical distribution of airborne LiDAR returns from forest stands.
We studied 5,428 regular, even-aged stands in Quebec (Canada) with five dominant species: balsam fir (\emph{Abies balsamea} (L.) Mill.), paper birch (\emph{Betula papyrifera} Marsh), black spruce (\emph{Picea mariana} (Mill.) BSP), white spruce (\emph{Picea glauca} Moench) and aspen (\emph{Populus tremuloides} Michx.).
We modeled the vertical distribution against the three variables using a functional general linear model and a novel nonparametric graphical test of significance.
Results indicate that LiDAR returns from aspen stands had the most uniform vertical distribution.
Balsam fir and white birch distributions were similar and centered at around 50\% of the stand height, and black spruce and white spruce distributions were skewed to below 30\% of stand height (\(p\)\textless0.001).
Increased crown cover concentrated the distributions around~50\% of stand height.
Increasing age gradually shifted the distributions higher in the stand for stands younger than 70-years, before plateauing and slowly declining at 90--120 years.
Results suggest that the vertical distributions of LiDAR returns depend on the three variables studied.

\hypertarget{introduction}{%
\section{Introduction}\label{introduction}}

The distribution of vegetation within canopies varies with tree
allometry and competition strategies, leading to variations in canopy
structure and ultimately, in environmental conditions
(Purves et al. \protect\hyperlink{ref-purves_etal07}{2007}; Thorpe et al. \protect\hyperlink{ref-thorpe_etal10}{2010}; Pretzsch and Dieler \protect\hyperlink{ref-pretzsch_dieler12}{2012}). Species‑specific canopy
structures create different
microhabitats, light conditions, and microclimates, which in turn
influence the rates at which stands sequester carbon. Species have different
carbon allocation strategies that evolve during the growth season,
relative to above and below-ground carbon allocation. Examples include growing fruit, deploying leaves, and growing roots for better access to nutrients and water
(De Pury and Farquhar \protect\hyperlink{ref-depury_farquhar97}{1997}; Lacointe \protect\hyperlink{ref-lacointe00}{2000}; Stark et al. \protect\hyperlink{ref-stark_etal12}{2012}). These growth and allocation strategies result in distinct species-specific tree shapes.

Traditionally, the characterization of stand vertical distribution of aboveground biomass required either on-site estimation of biomass per vertical layer (often requiring destructive measurements), scaffolding, or tree climbing, all of which are time-consuming and limit surveys to small areas (MacArthur and Horn \protect\hyperlink{ref-macarthur_horn69}{1969}; Aber \protect\hyperlink{ref-aber79}{1979}; Bassow and Bazzaz \protect\hyperlink{ref-bassow_bazzaz97}{1997}; Tackenberg \protect\hyperlink{ref-tackenberg07}{2007}).

The increasing availability of data collected remotely from airborne,
spaceborne and terrestrial sensors has improved our understanding of
forest dynamics (White et al. \protect\hyperlink{ref-white_etal13}{2013}; Wulder et al. \protect\hyperlink{ref-wulder_etal12}{2012}; Beland et al. \protect\hyperlink{ref-beland_etal19}{2019}).
Among these sensors, light detection and ranging
(LiDAR) has the ability to penetrate the canopy and provide information
on the spatial location of reflective material. LiDAR is increasingly
used to perform both extensive and highly detailed imaging of forest.
It uses a laser, at a precisely known location and orientation, to record a 3D scene.
A discrete LiDAR pulse is reflected back to the
sensor by the vegetation as it penetrates the canopy. The sensor then records
the total travel time of the pulse and its energy to deduce the distance from the
sensor to the location of reflection. LiDAR can be used from both the ground, and an aircraft to provide different perspectives on the forest. Terrestrial LiDAR provides highly detailed structural information about individual trees, but its extent is limited to a few hundreds of meters (Beland et al. \protect\hyperlink{ref-beland_etal19}{2019}; Crespo-Peremarch et al. \protect\hyperlink{ref-crespo-peremarch_etal20}{2020}).
Airborne LiDAR, on the other hand, is typically less detailed than terrestrial LiDAR but can be used for extensive landscape measurements.

Airborne LiDAR has been used to characterize vertical canopy structure, crown
shape, and aboveground biomass for entire ecosystems (Lefsky et al. \protect\hyperlink{ref-lefsky_etal02}{2002}; Parker et al. \protect\hyperlink{ref-parker_etal04}{2004}; Coops et al. \protect\hyperlink{ref-coops_etal07}{2007}; Stark et al. \protect\hyperlink{ref-stark_etal12}{2012}; Harding et al. \protect\hyperlink{ref-harding_etal01}{2001}; Cao et al. \protect\hyperlink{ref-cao_etal14}{2014}; Ellsworth and Reich \protect\hyperlink{ref-ellsworth_reich93}{1993}; Papa et al. \protect\hyperlink{ref-papa_etal20}{2020}). It has also been used to identify tree species (Heinzel and Koch \protect\hyperlink{ref-heinzel_koch11}{2011}, \protect\hyperlink{ref-heinzel_koch12}{2012}; Vaughn et al. \protect\hyperlink{ref-vaughn_etal12}{2012}; Axelsson et al. \protect\hyperlink{ref-axelsson_etal18}{2018}; Hovi et al. \protect\hyperlink{ref-hovi_etal16}{2016}; Fassnacht et al. \protect\hyperlink{ref-fassnacht_etal16}{2016}; Budei et al. \protect\hyperlink{ref-budei_etal18}{2018}; Fedrigo et al. \protect\hyperlink{ref-fedrigo_etal18}{2018}) and to study stand characteristics such as
age, crown cover, and basal area (Korhonen et al. \protect\hyperlink{ref-korhonen_etal11}{2011}; Racine et al. \protect\hyperlink{ref-racine_etal14}{2014}; White et al. \protect\hyperlink{ref-white_etal13}{2013}; Karna et al. \protect\hyperlink{ref-karna_etal19}{2019}).
Airborne LiDAR is an important tool for biomass quantification, and offers the opportunity to explore large-scale phenomena that could only be observed on the field (Vierling et al. \protect\hyperlink{ref-vierling_etal08}{2008}, \protect\hyperlink{ref-vierling_etal10}{2010}; Seavy et al. \protect\hyperlink{ref-seavy_etal09}{2009}; Karna et al. \protect\hyperlink{ref-karna_etal20}{2020}).

One limitation of the airborne LiDAR data is its inability to distinguish differences in foliage condition or in species spectral variation.
This is because it lacks the spectral information commonly used to classify species such as
variations in red, green, and blue or near-infrared multispectral imagery.
In response, the most common strategy for distinguishing species using LiDAR has been to differentiate individual tree shapes and texture (Holmgren and Persson \protect\hyperlink{ref-holmgren_persson04}{2004}; Kim et al. \protect\hyperlink{ref-kim_etal11}{2011}; Fassnacht et al. \protect\hyperlink{ref-fassnacht_etal16}{2016}), and add spectral information from multispectral imagery.
More recently multispectral LiDAR has also been used to distinguish stand species (Budei et al. \protect\hyperlink{ref-budei_etal18}{2018}; Budei and St-Onge \protect\hyperlink{ref-budei_st-onge18}{2018}).

Most studies on species classification using aerial LiDAR have focused on
species identification based on individual tree crowns.
However, tree crown delineation requires a large number of LiDAR returns and highly accurate registration of ground observations (Ørka et al. \protect\hyperlink{ref-orka_etal09}{2009}; Muss et al. \protect\hyperlink{ref-muss_etal11}{2011}).
Using a small observation area such as an individual tree crown causes a loss in the shape of the vertical distribution of LiDAR returns.
This is due to a decrease in the number of LiDAR returns, causing the distribution to become a collection of random variates.
Thus, although the accuracy of species identification would be improved by a very high point density in excess of 10 pt/m² (Fassnacht et al. \protect\hyperlink{ref-fassnacht_etal16}{2016}), it is difficult to apply and validate these approaches for use over large areas and with less-dense LiDAR datasets.

An alternative to individual tree-crown extraction approaches is the application of
area-based approaches (see e.g.~White et al. (\protect\hyperlink{ref-white_etal13}{2013})).
This approach generally uses a pixel or a stand on
which LiDAR returns are aggregated and predictors are derived before being used in a model to predict forest attributes.
Predictors are often derived from the vertical distribution of LiDAR returns: the number of LiDAR returns per height slice.
The vertical distribution of LiDAR returns is often
presented as quantiles, projections of quantiles (such as principal component analysis), or parametric functions (such as a Fourier, beta or Weibull
functions), which are used to predict stand attributes (Mehtätalo \protect\hyperlink{ref-mehtatalo06}{2006}; Coops et al. \protect\hyperlink{ref-coops_etal07}{2007}; Racine et al. \protect\hyperlink{ref-racine_etal14}{2014}; Maltamo et al. \protect\hyperlink{ref-maltamo_etal05}{2005}; Palace et al. \protect\hyperlink{ref-palace_etal15}{2015}; Magnussen et al. \protect\hyperlink{ref-magnussen_etal99}{1999}; Riggins et al. \protect\hyperlink{ref-riggins_etal09}{2009}; Falkowski et al. \protect\hyperlink{ref-falkowski_etal09}{2009}). The area-based method can be effective with
a point density as low as 1~pt/m², which reduces cost and require less processing compared to
tree-crown approaches (White et al. \protect\hyperlink{ref-white_etal13}{2013}).

Lowering the minimum LiDAR point density threshold for species classification
would increase the number of potential surveys where this method could be applied, and at the same time reduce the cost of acquisition. The species information could then be used for
extensive forest management, or landscape-scale studies. One way to
achieve area-based species mapping is to increase our understanding of the interaction
between LiDAR and stand-level vegetation. However, the vertical distribution of LiDAR
returns is difficult to analyze without resorting to dimension reduction, a method
that generally limits the interpretability of results.

We hypothesize that using
functional general linear models (GLM) and a novel non-parametric graphical
test of significance (Mrkvička et al. \protect\hyperlink{ref-mrkvicka_etal19}{2019}) would allow us to link the forest
attributes to the vertical distribution of returns from low-density LiDAR surveys. The test provides a framework to compare a function (i.e.~the vertical distribution of LiDAR returns) against categorical and continuous variables, as well as a visual understanding of the effects of the variables on the function.
Versions of non-parametric graphical tests have been applied to
economical data (Mrkvička et al.~2020), and non-parametric inference is
commonly used in neuroimaging (Winkler et al., 2014).
However, to our knowledge, our study is the first to use a functional GLM
combined with a non-parametric graphical test of significance in the field of ecology or remote sensing.

The information contained in the vertical distribution of LiDAR returns is
closely related to the distribution of the vegetation. Reduced crown cover (the proportion of area covered by vegetation; Gonsamo et al. \protect\hyperlink{ref-gonsamo_etal13}{2013}) typically increased the probability for the LiDAR returns to reach lower vegetation (Hilker et al. \protect\hyperlink{ref-hilker_etal10}{2010}), while age influences the vertical position of the crown within a stand (Coops et al. \protect\hyperlink{ref-coops_etal09}{2009}; Racine et al. \protect\hyperlink{ref-racine_etal14}{2014}).
Using direct foliage measurements, Aber (\protect\hyperlink{ref-aber79}{1979}) observed that the vertical concentration of the foliage evolved with stand age, and that the end point of forest succession seemed to reach an
equilibrium where the foliage was relatively evenly distributed within the
canopy. Martin-Ducup et al. (\protect\hyperlink{ref-martin-ducup_etal16}{2016}) noted that crown cover and stand
maturity affected the shapes of the crown of sugar maples when measured using
terrestrial LiDAR.

The shapes of different species influence the distribution of LiDAR
returns, but the interpretations from most studies are limited and hard to generalize across ecosystems or LiDAR surveys (Fassnacht et al. \protect\hyperlink{ref-fassnacht_etal16}{2016}). Some studies
explicitly compared the vertical distribution of LiDAR returns between species.
For example, Ørka et al. (\protect\hyperlink{ref-orka_etal09}{2009}) found that the first and last returns were more
dispersed in Birch (\emph{Betula} spp.) stands than in Norway spruce (\emph{Picea abies} (L.) Karst.).
The vertical distribution of first returns were also skewed toward the top of the canopy, and increased stand height was shown to influence the overall return distribution (Ørka et al. \protect\hyperlink{ref-orka_etal09}{2009}).

In this study, we verify that we can differentiate the distinct patterns
for individual species in the vertical distribution of LiDAR returns from a low-density
area-based survey. We hypothesize that the use of a functional GLM combined with a non-parametric graphical
test of significance makes it possible to identify these inter-species
differences in the vertical distribution patterns
after stand crown cover and age effects are accounted for.

\hypertarget{methods}{%
\section{Methods}\label{methods}}

\hypertarget{study-area}{%
\subsection{Study area}\label{study-area}}

\begin{figure}
\includegraphics[width=1\linewidth]{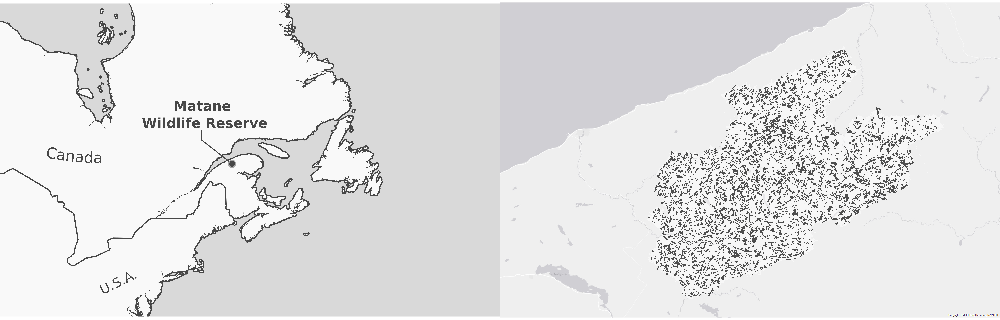} \caption[Location of Matane Wildlife Reserve and selected stands.]{Location of Matane Wildlife Reserve (left panel) and selected stands within the study area (right panel, dark patches).}\label{fig:fig-loc}
\end{figure}

The study was conducted in Matane Wildlife Reserve (Quebec, Canada,
48°41'N, 66°58'W) and covering 1,600 km² (Figure \ref{fig:fig-loc}). The reserve is a mixed
forest dominated by balsam fir (\emph{Abies balsamea} (L.) Mill.),
paper birch (\emph{Betula papyrifera} Marsh), and black spruce (\emph{Picea
mariana} (Mill.) BSP). Other species in the reserve include white
spruce (\emph{Picea glauca} Moench), aspen (\emph{Populus tremuloides }Michx.),
Norway spruce (\emph{Picea abies} (L.) Karst.), jack pine (\emph{Pinus banksiana}
Lamb.), and other non-commercial deciduous species. Most of the reserve is
subject to active commercial logging. Logging activities in that area have been documented since 1962; 27\% of
the area has been harvested (mostly total harvesting) and half of
that has been replanted.
Two percent of the stands originated from natural perturbations (e.g.~windthrow, insect epidemic, fire) and the remaining 71\% was undocumented.

\hypertarget{data}{%
\subsection{Data}\label{data}}

Airborne imagery and LiDAR were acquired during the summer of 2007. All
data were collected by the Quebec Ministry of Natural Resources as part
of their decennial forest mapping program.
The LiDAR survey used a nominal point density of 3 points/m² with an Optech ALTM 2050 sensor that recorded the first and last returns at 40kHz.
The survey was flown at 1,200 m above ground, with a flight overlap of 30\% and a maximal scan angle of 15° from nadir and a footprint diameter of 25~cm.

Airborne photography was conducted using a Leica ADS--40 pushbroom camera
at a resolution of 0.2~m for panchromatic, and 0.5~m for near-infrared, green, and blue bands (NIR, G, B respectively centered on 860, 560, and 460~nm); the side overlap was 40\% to ensure complete coverage of the area. This imagery was used to partition the territory into stands following provincial guidelines adapted from MRNQ (\protect\hyperlink{ref-mrnq07}{2007}) by expert photo-interpreters. Using digital
stereopsis (virtual 3D vision) from the forward and
nadir-facing ADS--40 sensor images, the expert photo-interpreters identified
tree species by integrating information on landscape positions, crown
shapes, textures, and colors. Every photo was segmented into homogeneous
stands using species, crown cover, height, and geomorphic criteria.
Each stand was classified based on species composition, crown cover,
age, height, and other ecological variables using a combination of
photography, ground-control points and historical data, which was used as a reference.
Stand age was estimated using ground control plots where trees were
cored. This information was then combined with available archives, height,
and ecology to estimate age from aerial photography. Stand age was
divided in six regular classes: 10 (0--20), 30 (21--40), 50
(41--60), 70 (61--80), 90 (81--100), and 120 (101,~∞), and irregular age classes (such as uneven-aged and multi-stratum stands).
Crown cover was also estimated by visually comparing the proportion of open
ground with the space occupied by the mature tree crowns. Crown covert was categorized in 9 classes: 10 (5--14), 20 (15--24), 30 (25--34), 40
(35--44), 50 (45--54), 60 (55--64), 70 (65--74), 80 (75--84), and
90 (85--100). Species were identified by their distinctive features in the composite images of near-infrared, green and blue
(NIR+G+B) mapped onto red, green, and blue (R, G, B)
channels, and their frequent associations in forest stands (Table \ref{tab:species}).
Finally, the interpretation of the aerial images relied on the ground-control points and the ecological knowledge of the photo-interpreters. We used this photo-interpreted forest map
as our reference data for species, crown cover, and age.

\begin{longtabu} to \linewidth {>{}l>{\raggedright}X>{\raggedright}X>{\raggedright}X}
\caption{\label{tab:species}Description of tree shapes, associated species, preferred conditions, and colors}\\
\toprule
Species & Silhouette & Description\textsuperscript{1, 2} & Associated species\textsuperscript{2}\\
\midrule
\em{Abies balsamea} & \includegraphics[valign=T,scale=0.5,raise=2mm]{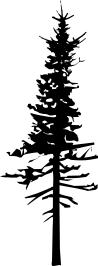} & Mesic sites. Narrow conic crown with a sharp and thin summit that is frequently pale due to accumulated cones that are more reflective than foliage. Brown, slightly pink tint. Browner and pinker than white spruce. & Trembling aspen, white birch, white spruce, black spruce, red spruce, and eastern hemlock\\
\cmidrule{1-4}
\em{Betula papyrifera} & \includegraphics[valign=T,scale=0.5,raise=2mm]{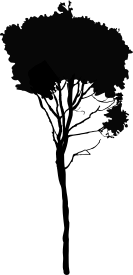} & Avoids sites with poor drainage. Flat half-sphere shape. Crown is  highly mingled with neighbors and exhibits an irregular texture that makes it hard to identify individual crowns even at higher resolutions. Dark pink tint between yellow birch and maples. & Various species including other birches, pines, spruces, hemlocks, poplars, maples, balsam fir, northern red oak, and pin cherry\\
\cmidrule{1-4}
\em{Picea mariana} & \includegraphics[valign=T,scale=0.5,raise=2mm]{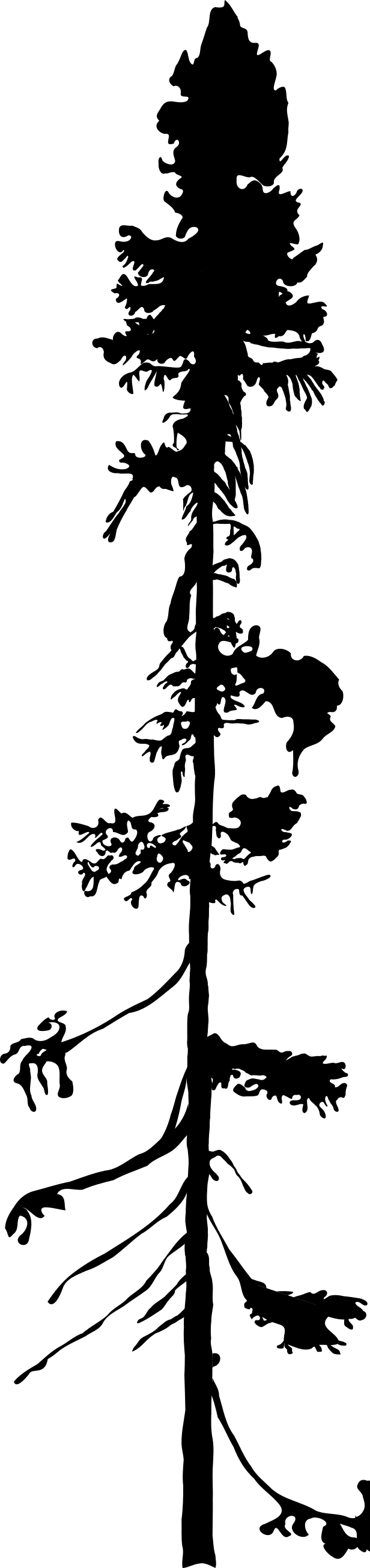} & Sites with poor drainage. Crown is thin and narrow, spirelike, with sharp summit and compact foliage. When located on well-drained upland sites, principal branches are shorter than other spruces. Lower branches droop and tips are upturned. Upper part of the crown is often very dense and oddly shaped with many cones. Brown, slightly pink when young. Paler than white spruce. & Tamarack in the southern part of the range. Jack pine, white spruce, balsam fir, white birch, trembling aspen in the northern part of the range\\
\cmidrule{1-4}
\em{Picea glauca} & \includegraphics[valign=T,scale=0.5,raise=2mm]{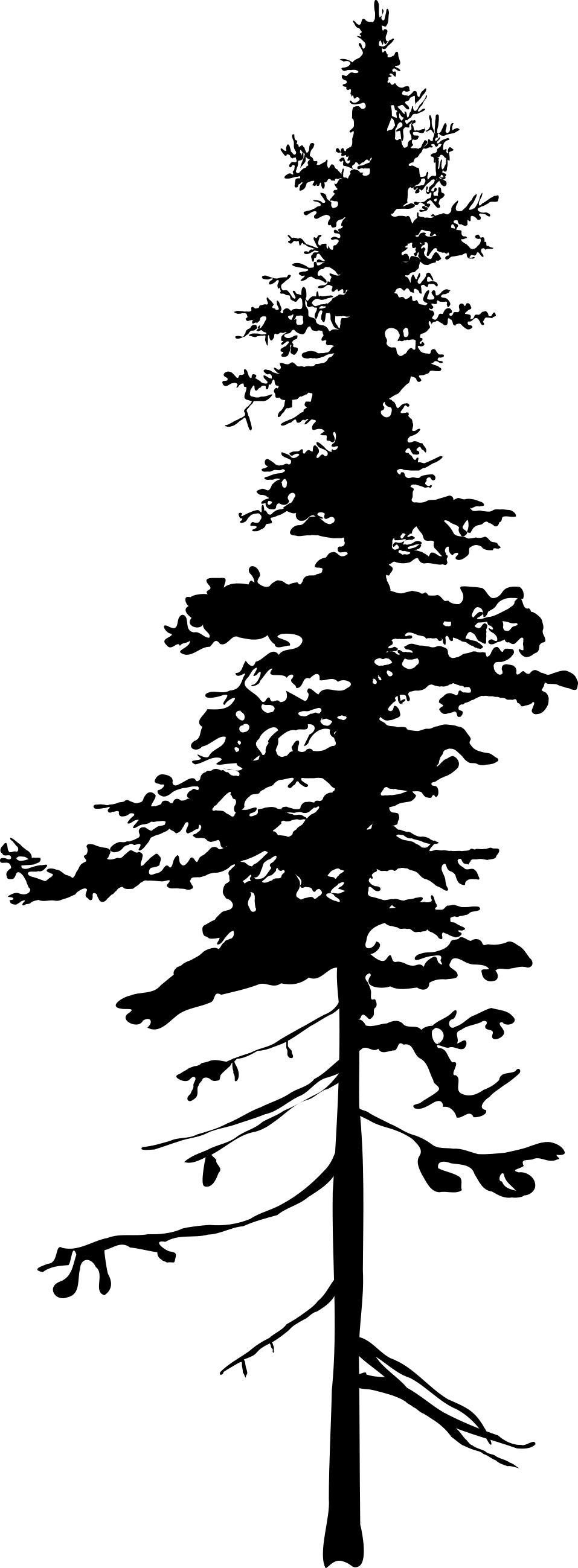} & Mid slope sites with good to moderate drainage. Broad conical crown is star-shaped, ragged, irregular, densely foliated, and spirelike in northern parts of its range. Principal branches are bushy, generally horizontal, and sometimes sloping downward in the lower part of the crown, with tips gradually upturned. Brown, slightly pink when young. Darker than black spruces. & Trembling aspen, white birch, black spruce, and balsam fir\\
\cmidrule{1-4}
\em{Populus tremuloides} & \includegraphics[valign=T,scale=0.5,raise=2mm]{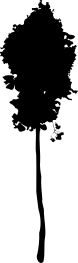} & All sites except those with poor drainage. Short, rounded, light bulb-shaped crown. Usually taller than surroundings when mixed. Crown surface looks blurred and smooth because of its small leaves. Orange-tinted pink. & White spruce, black spruce, balsam fir, white birch, balsam poplar and jack pine\\
\bottomrule
\multicolumn{4}{l}{\rule{0pt}{1em}\textsuperscript{1} Colors are based on (NIR+G+B mapped to R, G, B channels).}\\
\multicolumn{4}{l}{\rule{0pt}{1em}\textsuperscript{2} Adapted from Farrar (1995); Leboeuf and Vaillancourt (2013a, 2013b)}\\
\end{longtabu}

Even-aged stands are less complex than irregular stands when analyzed from a vertical structure point of view.
Therefore, we focused our analysis on even-aged stands where clearly dominant species represented at least 50\% of the stand cover.
Using a low threshold for species dominance increased the number of usable observations for the analysis.
This improved the ability of the analysis to detect effects, but also increased the noise from other co‑dominant species.

\hypertarget{lidar-data-processing}{%
\subsection{LiDAR Data Processing}\label{lidar-data-processing}}

\begin{figure}
\includegraphics[width=1\linewidth]{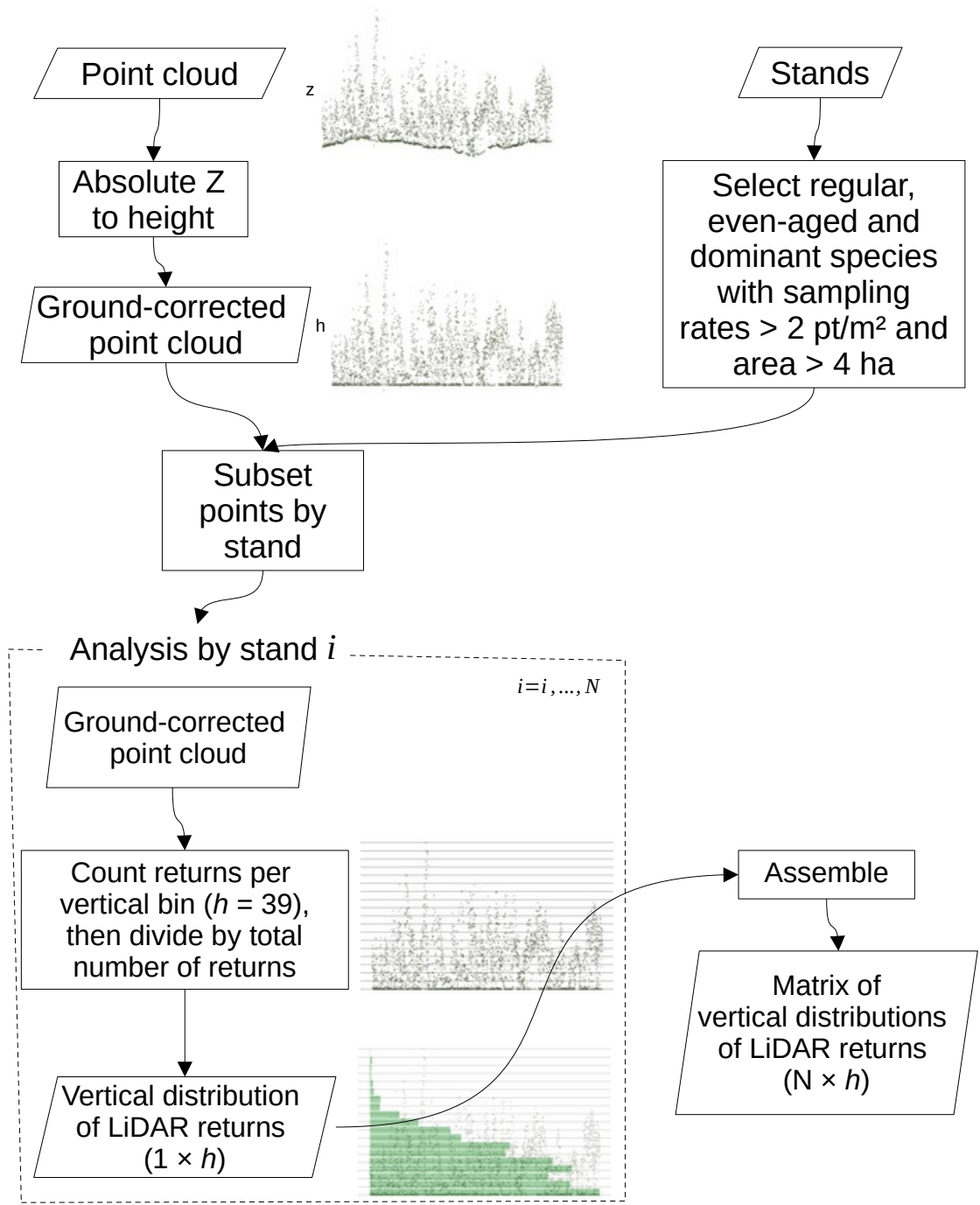} \caption[Preparation of the LiDAR data]{Step-by-step preparation of the LiDAR data}\label{fig:fig-lidar-process}
\end{figure}

The steps required to prepare LiDAR data for processing
are summarized in Figure \ref{fig:fig-lidar-process}.
We excluded stands with average LiDAR sampling
rates below 2~pt/m² or an area of less than 4~ha to ensure a sufficient
number of LiDAR returns. From the remaining stands, we kept only those with a dominant species that was observed at least in 100 stands: aspen,
balsam fir, black spruce, paper birch and white spruce. From the 22,365
stands on the original forest map, we reduced our dataset to
5,428 stands representing 35\% of the study area (Figure \ref{fig:fig-loc}).
Stand area distributions were similar for all selected species (median of 9~ha, minimum of 4~ha, maximum of 137~ha).

We registered the LiDAR returns by scaling their height between 0 and 1 to allow comparisons across stands of different heights.
We subtracted the ground elevation from the absolute point cloud elevation (Muss et al. \protect\hyperlink{ref-muss_etal11}{2011}).
For each stand, LiDAR returns were binned into a vertical distribution histogram of 39 height slices from their highest LiDAR return (Coops et al. \protect\hyperlink{ref-coops_etal07}{2007}; Harding et al. \protect\hyperlink{ref-harding_etal01}{2001}).
We then divided each slice count by the total number of points in the stand, so the vertical distribution of each stand summed to one, regardless of their inconsistent shapes, areas, and LiDAR point densities.

\hypertarget{statistical-analysis}{%
\subsection{Statistical Analysis}\label{statistical-analysis}}

We used a novel method that builds on the functional data analysis field
(Ramsay and Silverman \protect\hyperlink{ref-ramsay_silverman05}{2005}) to compare the complete vertical distribution of LiDAR
returns. Most studies of species classification have focused on the accuracy of classifiers and the value of predictors for species identification (Fassnacht et al. \protect\hyperlink{ref-fassnacht_etal16}{2016}), often using
dimension reduction to decrease and decorrelate the number of predictors.
Some of these reduction methods include linear discriminant analysis and principal component analysis (Koenig and Höfle \protect\hyperlink{ref-koenig_hofle16}{2016}; Räty et al. \protect\hyperlink{ref-raty_etal16}{2016}; Axelsson et al. \protect\hyperlink{ref-axelsson_etal18}{2018}). However, dimension reduction diminishes the
ability to understand the effect of individual variables and interpret the results of the model.

To compare inter- and intraspecific variations in the distribution of LiDAR returns and the
effect of three variables, we used a nonparametric graphical test of
significance (Mrkvička et al. \protect\hyperlink{ref-mrkvicka_etal19}{2019}; Myllymäki and Mrkvička \protect\hyperlink{ref-myllymaki_mrkvicka20}{2020}).
We modeled the vertical
distributions of LiDAR returns as a function of dominant species, crown cover,
and age using the general linear model

\begin{equation}
d(h) = \beta_0(h) +X_{sp}\cdot\beta_{sp}(h) +X_{cc}\cdot\beta_{cc}(h) +X_{age}\cdot\beta_{age}(h) +\varepsilon(h)      \label{eq:full}
\end{equation}

where, for every scaled height \(h\), \(d(h)\) is the \(N\times1\) vector of observed LiDAR distributions at scaled height \(h\), and
\(\beta_{sp}(h)\), \(\beta_{cc}(h)\), and \(\beta_{age}(h)\) are the parameter vectors related to species in \(X_{sp}\), crown cover in \(X_{cc}\), and age in \(X_{age}\); \(\varepsilon(h)\) is the vector of random errors with mean zero and finite variance \(\sigma^2(h)\). The crown cover was considered a continuous variable, while species and age were considered categorical variables (given that the last age class was open). This model that incorporates all three variables is referred to as the full model.

\begin{figure}
\includegraphics[width=1\linewidth]{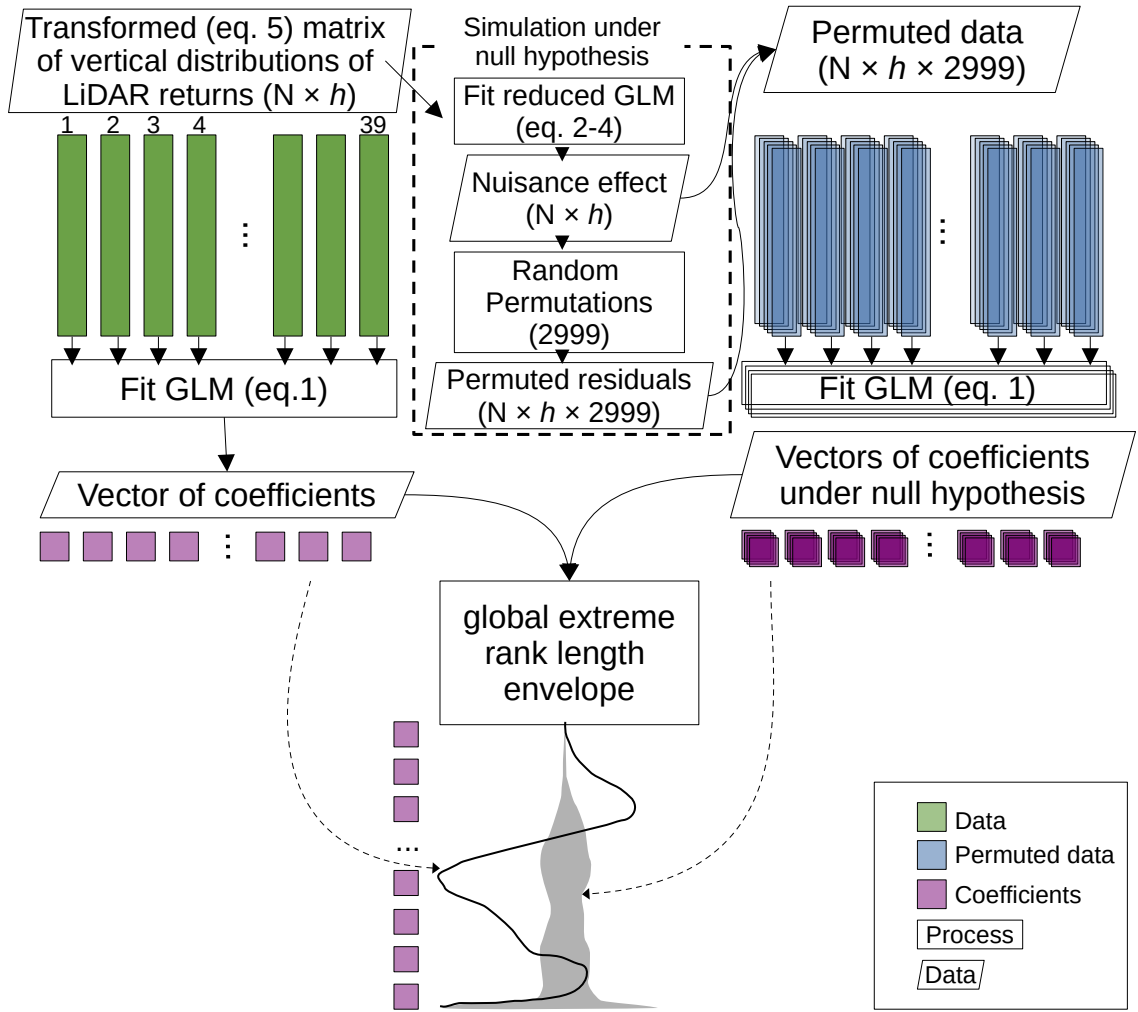} \caption[Description of the steps of the graphical test of significance]{Description of the steps of the graphical test of significance of a variable (e.g. species) on the vertical distribution of LiDAR returns.}\label{fig:fig-fglm-process}
\end{figure}

We studied the effect of each variable after accounting for the other
variables (also called the nuisance factors by Freedman and Lane
(\protect\hyperlink{ref-freedman_lane83}{1983})) using the methodology illustrated in Figure \ref{fig:fig-fglm-process}. We tested the effect of each three variables using the following null hypothesis: \(\beta_{sp, m}(h)=0\) for all \(m\) and \(h\), \(\beta_{cc}(h)=0\) for all \(h\), and \(\beta_{age, l}(h)=0\) for all \(l\) and \(h\), where \(m\) and \(l\) refer to the elements of the parameter vectors \(\beta_{sp}(h)\) and \(\beta_{age}(h)\) for each species and age groups, respectively.
The \(\beta\) coefficients of the discrete factors were constrained to sum to zero.
The corresponding three null models were obtained by removing the
studied variable from the full model (Eq. \eqref{eq:full}):

\begin{equation}
d(h) = \beta_0(h) +X_{cc}\cdot\beta_{cc}(h) +X_{age}\cdot\beta_{age}(h) +\varepsilon(h) \label{eq:species}
\end{equation}

\begin{equation}
d(h) = \beta_0(h) +X_{sp}\cdot\beta_{sp}(h) +X_{age}\cdot\beta_{age}(h) +\varepsilon(h)   \label{eq:cc}
\end{equation}

\begin{equation}
d(h) = \beta_0(h) +X_{sp}\cdot\beta_{sp}(h) +X_{cc}\cdot\beta_{cc}(h) +\varepsilon(h)     \label{eq:age}
\end{equation}

We used the coefficients of the variable of interest
(the left-out \(\beta\) in Equations \eqref{eq:species}--\eqref{eq:age}) for the
statistical test as suggested by Mrkvička et al. (\protect\hyperlink{ref-mrkvicka_etal19}{2019}).
For the continuous crown cover variable, the test statistic we used
was the vector \(\beta_{cc}(h)\) for all \(h\); and for age, the test was based on the
values of the effect \(\beta_{age,l}(h)\) of all the age groups \(l\) for all \(h\).
To examine species differences, the test was based on all differences
\(\beta_{sp, m}(h)-\beta_{sp, n}(h)\) for species \(m\) and \(n\) with \(1≤m<n≤5\).

The test we used relies on two procedures: 1) the Freedman-Lane algorithm (described in details by Winkler et al. \protect\hyperlink{ref-winkler_etal14}{2014}, p. 385) to permute the
residuals of the null model and create the reference distribution of the
coefficients under the null hypothesis, and 2) the global extreme rank length envelope test to build a null global envelope for the above-mentioned test
statistics and correct for the multiple tests conducted along \(h\)
(Myllymäki et al. \protect\hyperlink{ref-myllymaki_etal17}{2017}).
The Freedman-Lane algorithm includes all the steps ranging from the simulation under the null hypothesis to the final estimation of the chosen test statistic (Figure \ref{fig:fig-fglm-process}).
We used 2,999 random permutations to build the distributions under the null hypothesis. The global extreme rank length envelope test (Myllymäki et al., 2017, Mrkvička et al., 2020) was then constructed from the test statistics calculated from the empirical vertical distribution of LiDAR returns and the 2,999 permuted data sets.

The null hypothesis was rejected if the empirical test vector left the 98\% global envelope at any point. To account for the three variables tested (species, crown cover, and age), we used a Bonferroni adjusted significance level \(\alpha = 0.05/3\) in addition
to the inherent correction applied within the global extreme rank length
envelope test to account for multiple \(h\).
We observed that the variances of the model residuals were heterogeneous for all the variables (Winkler et al. \protect\hyperlink{ref-winkler_etal14}{2014}).
Following Mrkvička et al. (\protect\hyperlink{ref-mrkvicka_etal20}{2020}), we transformed the matrix of vertical LiDAR returns distribution (Figure \ref{fig:fig-fglm-process}) by scaling the functions \(d_{i}\) of each group \(j\) according to their variance dispersion. The initial \(d_{i,j}\) function was
transformed into a \(S_{i,j}\) (scaled) function by

\begin{equation}
S_{i,j}(h) = \frac{d_{i,j}(h) - \bar{d_j}(h)}{\sqrt{Var(d_j(h))}} \cdot\sqrt{Var(d(h))} + \bar{d_j}(h)       \label{eq:variance}
\end{equation}

where the group sample variance \(Var(d_j(h))\) is used to correct for unequal
variance among groups. The group sample mean \(\bar{d_j}(h)\) and overall variance \(Var(d(h))\)
are used to preserve the original scale of the mean and variability of the functions.

Our experiments with simulated data showed that transforming all groups at once (by combining all group levels) removed the heterogeneity of the variance. However, this required there to be sufficient observations in all categories, which was not the case for our data.
We therefore relied on the successive application of the transformation
(Equation \eqref{eq:variance}) for each of the three variables, and we found that the order in which the transformation is applied can reintroduce heterogeneity of variance. We
settled on the successive transformation of crown cover, age, and species
which provided the best results and reduced heterogeneity. We verified
the importance of the correlation by running Breusch-Pagan test of
heteroscedasticity (Breusch and Pagan \protect\hyperlink{ref-breusch_pagan79}{1979}) using the squared residuals \((d(h) - \hat{d}(h))²\) on
the left-hand side of the reduced equations \eqref{eq:species}--\eqref{eq:age}.
While the test indicated significant heteroscedasticity in some areas of the
curves, the coefficient of determination was less than 4\% for all variables and
all \(h\), which confirmed that no further adjustments were required.

We performed the analysis using R version 3.6.3 (R Core Team \protect\hyperlink{ref-rcoreteam20}{2020}), the GET package
(Mrkvička et al. \protect\hyperlink{ref-mrkvicka_etal19}{2019}; Myllymäki and Mrkvička \protect\hyperlink{ref-myllymaki_mrkvicka20}{2020}), and Lastools (Isenburg \protect\hyperlink{ref-isenburg12}{2012}) to correct
and extract the LiDAR data.

\hypertarget{results}{%
\section{Results}\label{results}}

\begin{figure}
\includegraphics[width=1\linewidth]{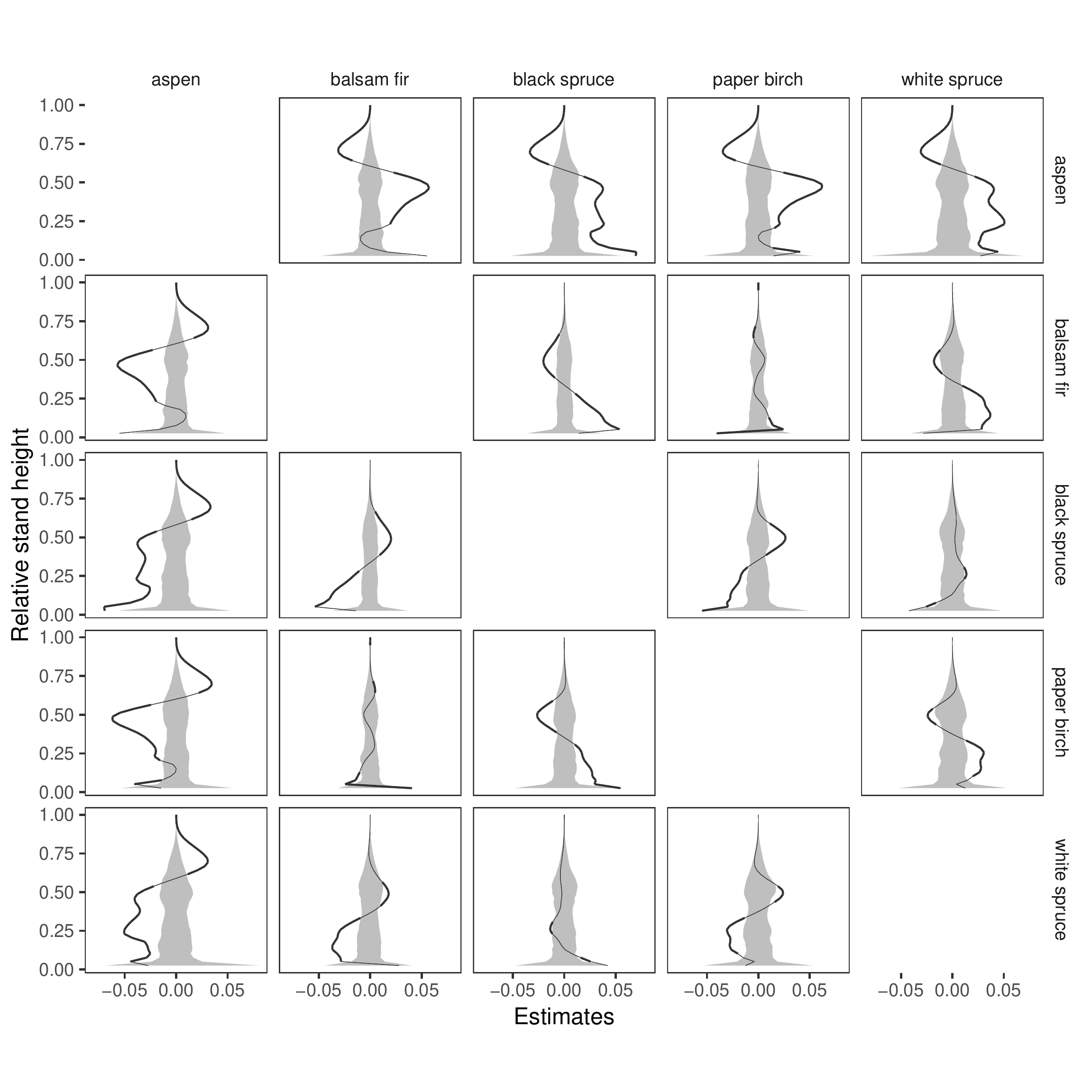} \caption[Nonparametric graphical tests of significance comparing species]{Nonparametric graphical tests of significance comparing species using contrasts: the observed difference between the coefficients of two species (black curve), where the species in the rows are subtracted from the species in the columns (e.g. first column, second row is the aspen − balsam fir contrast). The 98\% global envelope (grey bands), shows the area of acceptance of the null hypothesis (no effect, $p$<0.001) obtained from the permutations of the residuals of the null model (Equation \eqref{eq:species}). The observed curve that is outside the envelope is in bold. Panels above the diagonal are the reflection of the observed functions and the global envelope from the lower part (the tests were performed only once).}\label{fig:fig-sp}
\end{figure}

\begin{figure}
\includegraphics[width=1\linewidth]{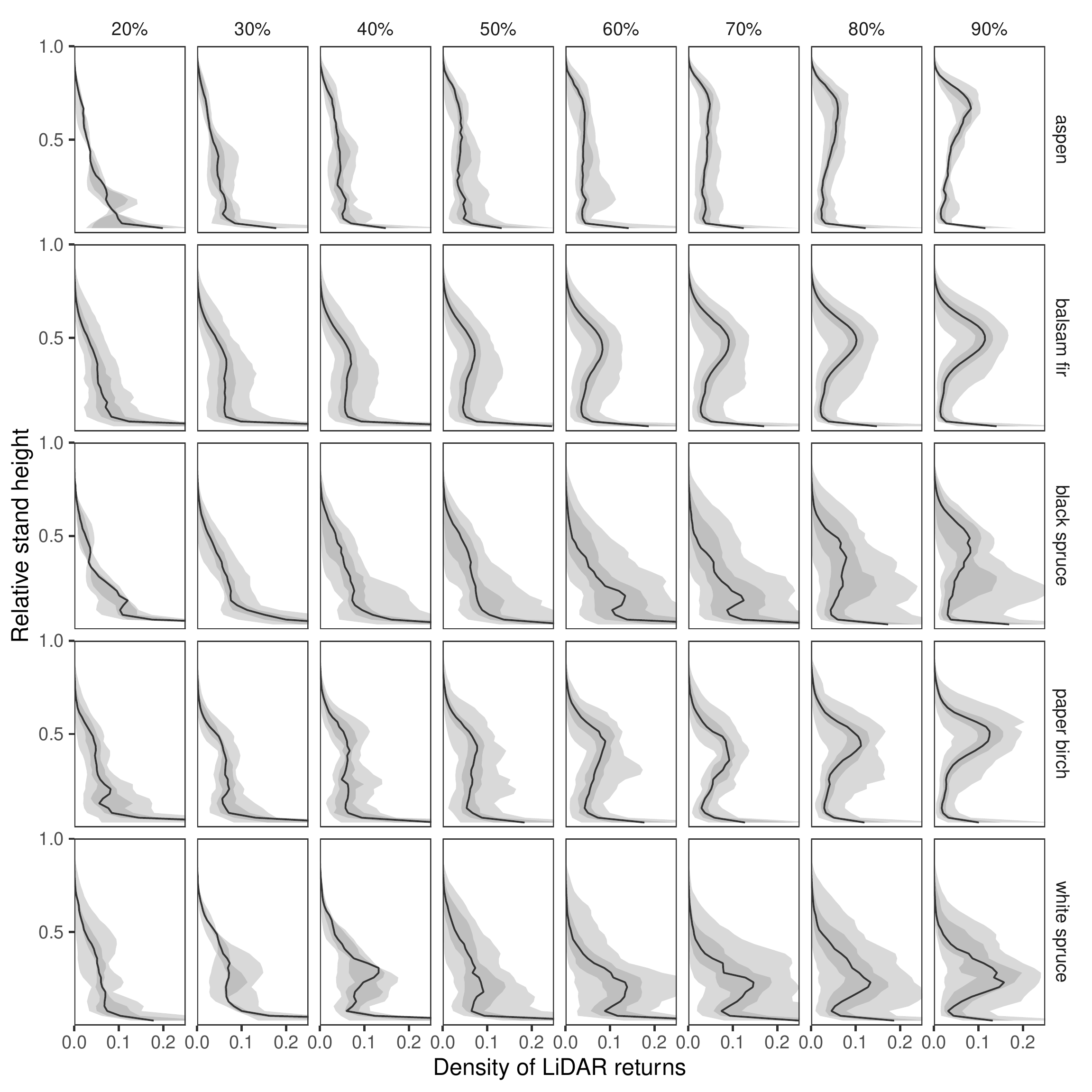} \caption[Vertical distribution of LiDAR returns as a function of crown cover and species.]{Vertical distribution of LiDAR returns as a function of crown cover (columns) and species (rows). Black lines represent the median distribution; shaded areas represent 95\% and 50\% local variation envelopes.}\label{fig:fig-cc2}
\end{figure}

\begin{figure}
\includegraphics[width=1\linewidth]{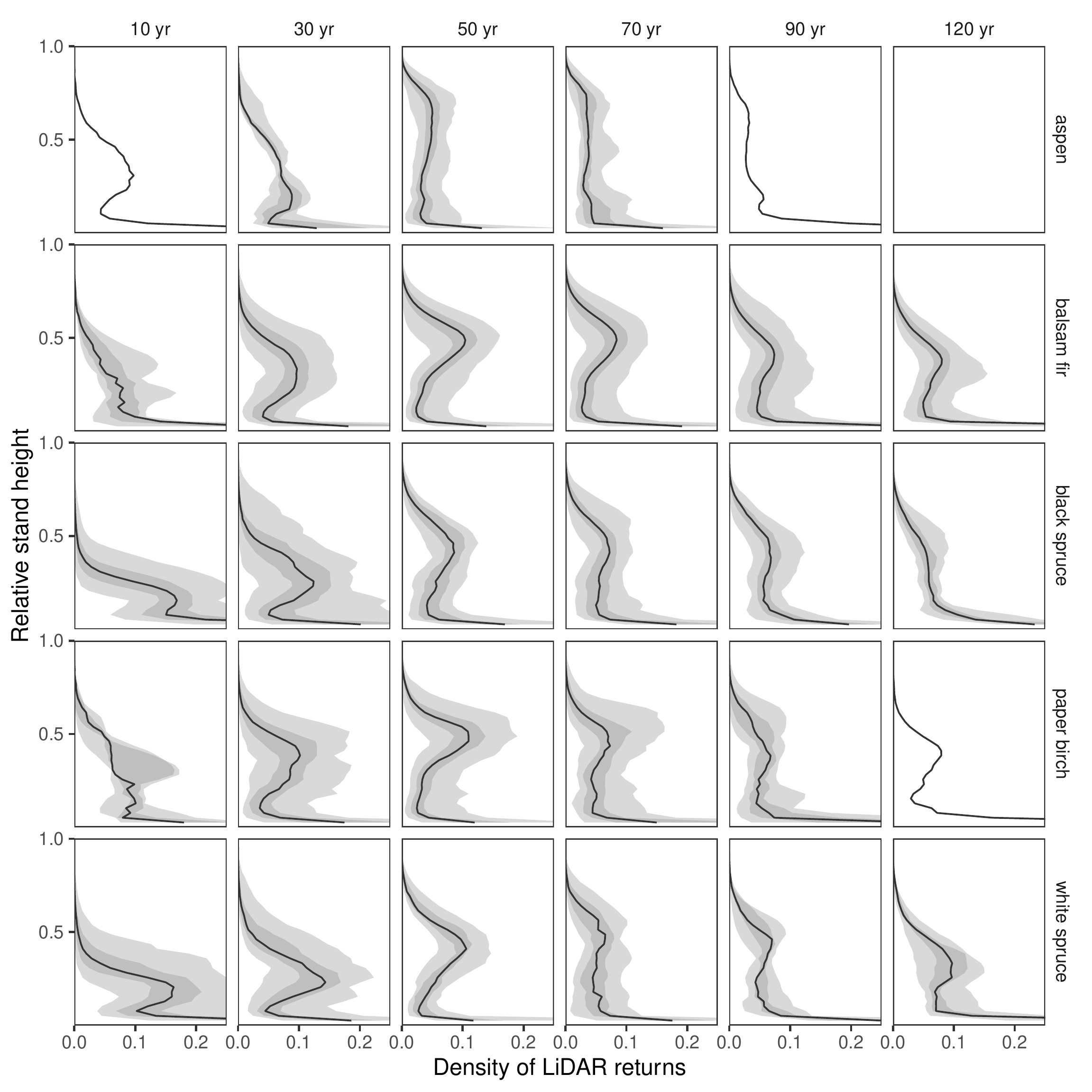} \caption[Vertical distribution of LiDAR returns as a function of age groups and species.]{Vertical distribution of LiDAR returns as a function of age (columns) and species (rows). Black lines represent the median distribution; shaded areas represent 95\% and 50\% local variation envelopes.}\label{fig:fig-age2}
\end{figure}

The comparisons from the nonparametric graphical test of significance
confirm that the differences between
all species, after accounting for age and crown cover, were significant (\(p\) \textless{} 0.001) (Figure \ref{fig:fig-sp}). We observed two groups of
species where differences were significant but small: balsam fir--paper
birch, and black spruce--white spruce. Differences between balsam fir and paper birch
were small and localized at 65--70\% and below 12\% of stand height. White spruce
and black spruce also had small localized differences at 22--30\% and
5--8\% of stand height. However, some differences between groups of
species were larger (indicated by the bold lines in Figure \ref{fig:fig-sp}): black and white spruces had the lowest distribution of LiDAR
returns when compared to other species. Black spruce had fewer returns
than balsam fir between 38--62\% of stand height (40--57\% for paper birch), while
it had more returns below 28\% of stand height (30\% for paper birch).
The vertical distribution of LiDAR returns for aspens had a distinctive shape when compared to every other species: it did not display a prominent peak,
which made the distribution more uniform than for other species
(Figures \ref{fig:fig-cc2} and \ref{fig:fig-age2}). Aspen was significantly different from other species (\(p\)\textless0.001) except in the
areas between 57--60\% of height, and 8--18\% for balsam
fir and paper birch specifically.
Overall, the areas with the largest differences between species
were located at around 5, 25, 50, and 70\% of stand height.

\begin{figure}
\includegraphics[width=1\linewidth]{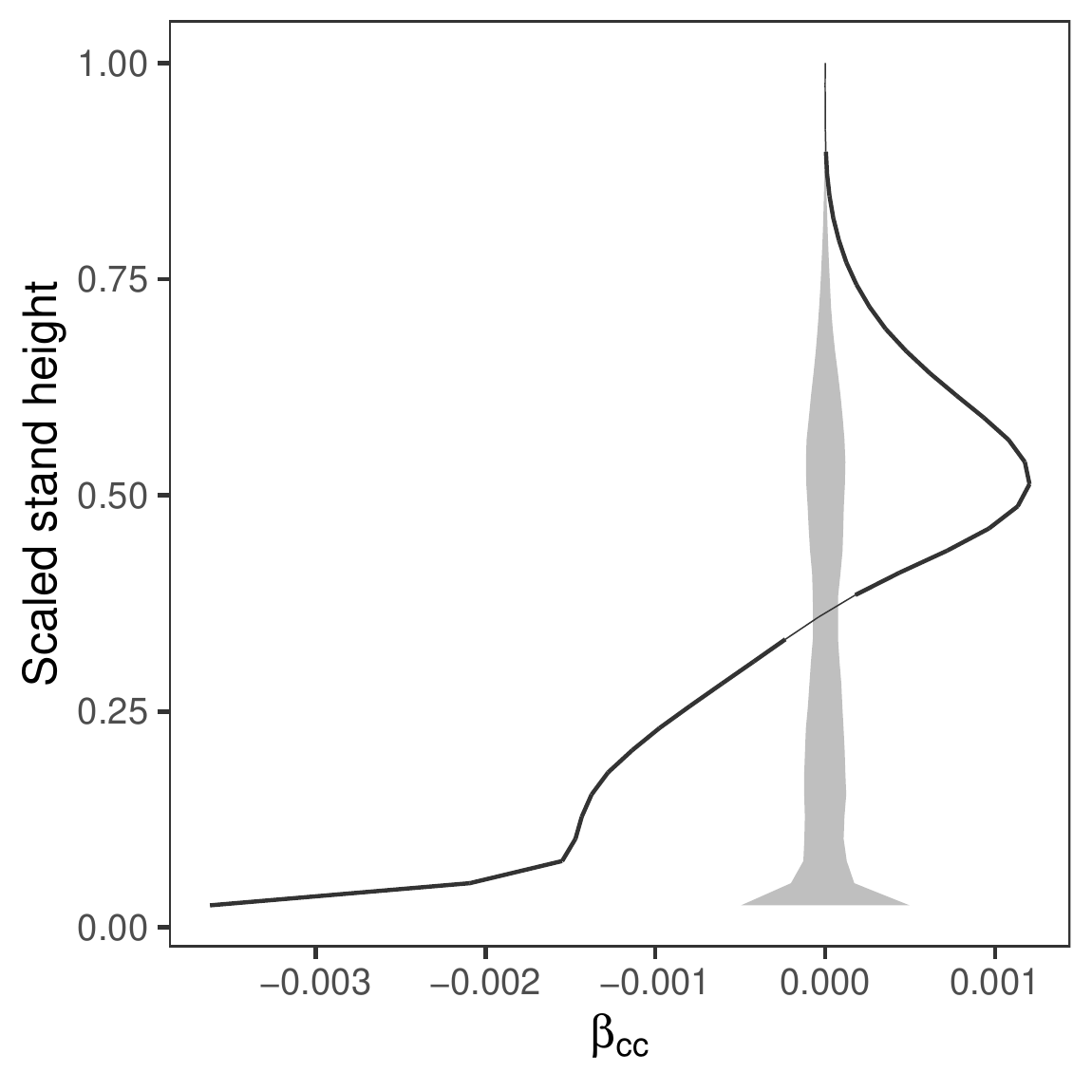} \caption[Nonparametric graphical tests of significance for crown cover]{Nonparametric graphical tests of significance for crown cover. The observed coefficient (black curve) and the 98\% global envelope (grey band) that shows the area of acceptance of the null hypothesis (no effect, $p$<0.001) obtained from permutations of the residuals of the null model (Equation \eqref{eq:cc}). The observed curve that is outside the envelope in in bold.}\label{fig:fig-cc}
\end{figure}

An increase in crown cover was associated with a decrease in the density of
LiDAR returns below 33\% of stand height, while the density of LiDAR returns
above 38\% increased (Figure \ref{fig:fig-cc}). The largest increase was
concentrated around the middle height (50\%), while the largest reduction effect occurred closer to the ground (around 3\% height).
Figure \ref{fig:fig-cc2} displays the vertical distribution of LiDAR returns for each species per crown cover (including the effect of age).
The increased crown cover concentrated the
returns at around 50\% of the stand height, except in the case of aspen, where the peak was higher in the stand (around 80\% of stand height), and white spruce, where the peak was lower (around 20\%).
Although the black spruce peak was centered at~50\% of the stand height for high values of crown cover, the observed
variability was skewed toward the lower stand heights. Some species were more widely
represented in younger or older age classes, which inflated the variation envelopes.
The model from Figure \ref{fig:fig-cc} accounted for this effect.

\begin{figure}
\includegraphics[width=1\linewidth]{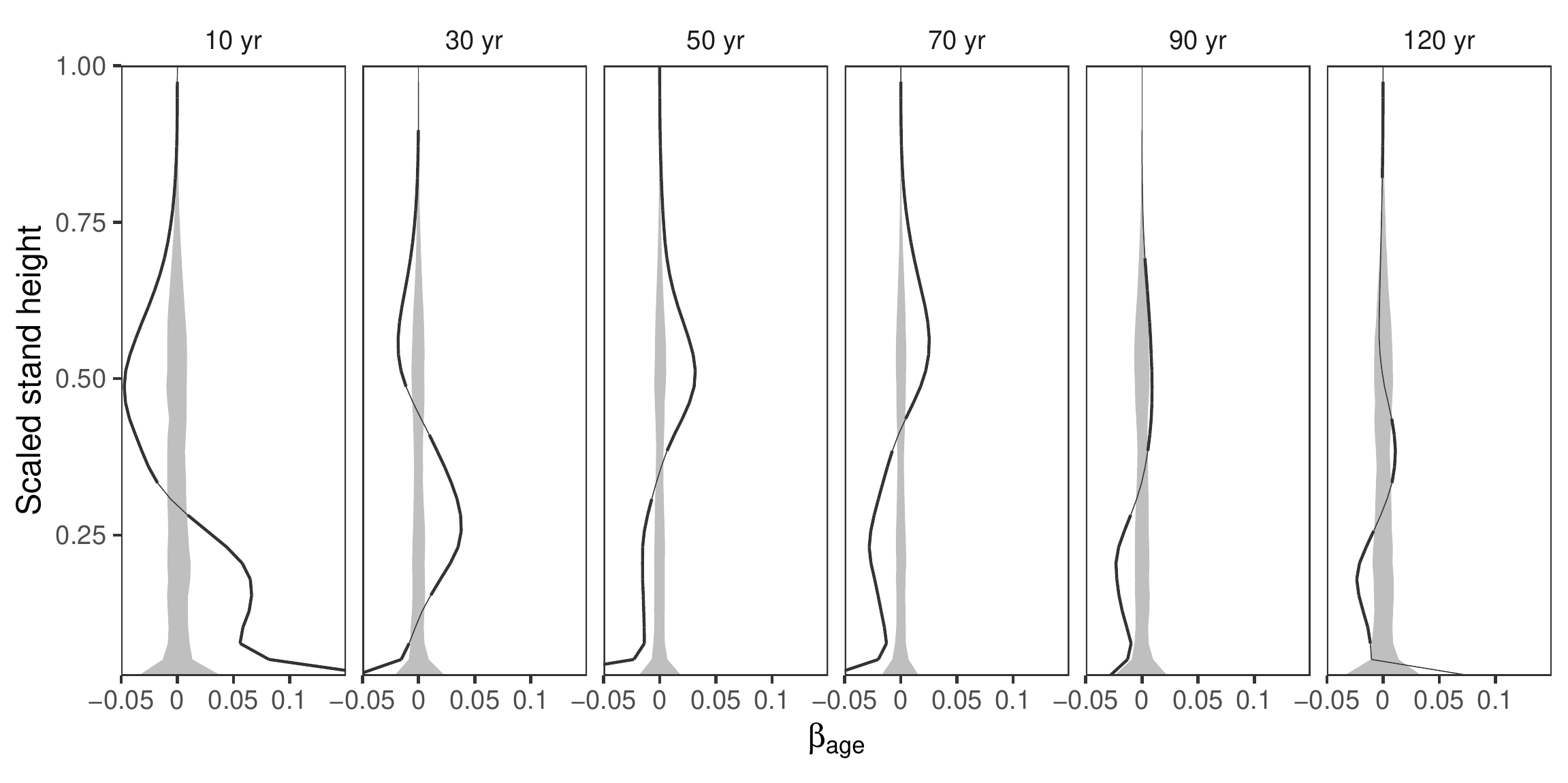} \caption[Nonparametric graphical tests of significance for age]{ Nonparametric graphical tests of significance for age. The observed coefficients of the six age groups (black curves), and the 98\% global envelope (grey bands below the diagonal) that shows the area of acceptance of the null hypothesis (no effect, $p$<0.001) obtained from the permutations of the residuals of the null model (Equation \eqref{eq:age}). The observed curve outside the envelope is in bold. }\label{fig:fig-age}
\end{figure}

Increased age was associated to a gradual upwards shift in LiDAR return distribution (Figure \ref{fig:fig-age}).
This shift occurred until 70 years of age, when a plateau was reached.
The effect of age was significant at most stand height between 10 and 70 years of age.\\
Stands that were in the 10- and 30-year age groups had an abundance of returns below 28\% and 41\% of stand height, respectively, and fewer returns above 33\% and 49\%, respectively.
However, the effect is reversed in stands in the 50- and 70- year groups, where age inflated the distribution between 38--97\% of stand height (44--100\% for 70~yr), and deflated it below 31\% of stand height (38\% for 70~yr).
The significant age effects for stands in the 90- and 120-year age groups were smaller, and occurred below 69\% and 44\% of the stand height, respectively (\(p\)\textless0.001).
These stands also had an abundance of returns in the middle and upper part of the stand (38--68\% for 90 yr; 33--44\% for 120 yr).
For stands that were in the 90- and 120-year groups, there were reduced numbers of returns below 28\% and 23\% of stand height, respectively.

Overall, the rate of change in the vertical distribution of LiDAR
returns gradually decreased as the age increased. For example, changes
in the distribution between 10 and 30 years were larger than the changes
between 70 and 90 years. Areas of higher variability for age
groups were located around 5\%, 20\%, and 50\% of stand height.
Figure \ref{fig:fig-age2} displays the median vertical distribution of LiDAR returns for each species, by age class (including the crown cover effect).
The young white spruce and black spruce stands displayed
a high degree of asymmetry toward the lower part of the stand that disappeared in
older stands.

\begin{figure}
\includegraphics[width=0.6\linewidth]{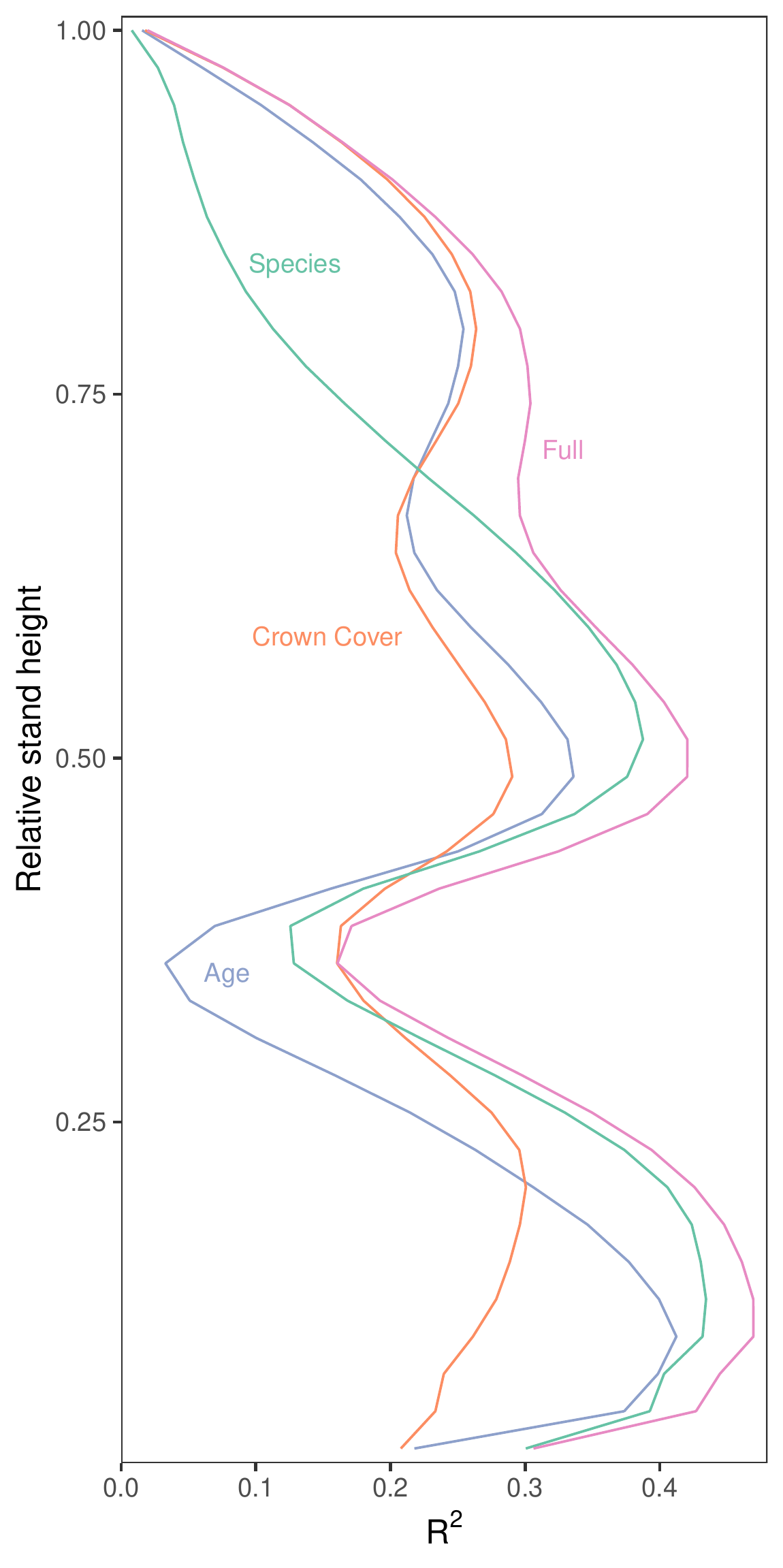} \caption[Comparison of $R²(h)$.]{Comparison of $R²(h)$ of the full model (Equation \eqref{eq:full}), and the three reduced models (Equations \eqref{eq:species}--\eqref{eq:age}) where one variable (either Species, Crown Cover, or Age) was excluded.}\label{fig:fig-rsq}
\end{figure}

The coefficient of determination for the models varied across stand heights (Figure
\ref{fig:fig-rsq}). The most correlated areas for every model were around
5--15\% and 46--54\% of stand height.
All models displayed a sharp decline between 28--40\%.
The highest R² was for the full model at 0.47 (at 10--13\% of stand height) and 0.42 (at 49--51\% of stand height).
The model that excluded the species variable (Equation \eqref{eq:species}) produced a difference in R² of more than 0.10 for 72--92\% stand height compared to the full model.
When the crown cover variable was excluded from the model (Equation \eqref{eq:cc}) for between 0--21\% and 46--67\% of stand height, there was a difference in R² of more than 0.10 with the full model.
The model that excluded the age variable (Equation \eqref{eq:age})
produced a difference in R² of more than 0.10 with the full model at 18--38\% and 62--64\% of stand height.

\hypertarget{discussion}{%
\section{Discussion}\label{discussion}}

Our objective was to study the influence of species, crown
cover and age on the vertical distribution of LiDAR returns. We found
that even-aged stands exhibit species-specific patterns that predictably evolve with crown cover and age. We observed two groups of
species: the first, balsam fir and paper birch had more symmetrical
vertical distributions of LiDAR returns that were centered between 40\% and 60\% of stand height, and the second, white spruce and black spruce, had distributions of LiDAR returns that were generally skewed lower in the stand (below 30\% of the stand height).
Aspen displayed a more even distribution of LiDAR returns with a higher proportion of returns higher in the stand compared to other species.

While individual trees are plastic and can adapt to various light and environmental conditions, we found that stands of the same species share similar vertical characteristics.
These characteristics distinguish them from other species.
This observation is consistent with previous field observations that were conducted over a smaller area using different measurement methods (Purves et al. \protect\hyperlink{ref-purves_etal07}{2007}).
However, we expected clearer vertical distribution patterns along shade-tolerance gradients or between conifer and deciduous trees; we found that patterns of paper birch were more similar to those of balsam fir (a shade-tolerant) than aspen, another shade-intolerant deciduous.
Since we used the dominant species to classify each stand, the similar vertical distribution patterns could be the result of frequent occurrence of balsam fir and paper birch within the same stand.
The effect of species associations could be controlled by using a higher dominance threshold than the 50\% we used.
However, we found that using a 60\% dominance made little difference to the conclusions, whereas a higher threshold removed too much data for some species.

The results in this paper demonstrate that vertical LiDAR return distribution can be useful for species classification.
Based on these findings, models could be trained to link these vertical distributions to species and then used to estimate species occurrence over the landscape.
Many studies performing species classification with LiDAR use the vertical distribution of LiDAR returns (Fassnacht et al. \protect\hyperlink{ref-fassnacht_etal16}{2016}), however the approach developed here allows the drivers of cover and age to also be integrated into the estimation, thereby making the species prediction more robust when applied more broadly.

Additional evidence is required in order to establish whether the relationships between vertical density of LiDAR returns and the species, crown cover, and age are specific to our study area.
Including other variables, such as LiDAR scan angle and abiotic factors, might improve the model, making it more applicable to other study areas or complex stand structures.

Increased crown cover led to an increased concentration of the vertical distribution of LiDAR returns to higher in the stand.
In turn, stands with decreased crown cover exhibited a vertical distribution of LiDAR returns that was more dispersed and lower to the ground, with fewer returns higher in the stand.
The gradual displacement of LiDAR returns from below 33\% to above 38\% associated to crown cover was also apparent when we compared the variance explained by the models in Figure \ref{fig:fig-rsq} where crown cover appeared to explain at least 25\% of the variance of the full model below 21\% and between 46--67\% of stand height.

Increased age was associated with the vertical distribution of LiDAR returns being displace higher in the stand for up to 70 years, followed by a plateau or decline in the 90- and 120-year-old groups.
The rate of change between age groups decreased as the age increased. Vertical distribution of LiDAR returns from younger stands appeared to undergo a rapid transformation in the 10-year group, which gradually decelerated and stabilized in the 70-year-old group.
The 90- and 120-year-old groups displayed the least amount of change.
While it is well established that absolute stand height varies predictably with age, the changes we observed here are relative to the maximal height of the stand.

The effect of age was akin to that of crown cover: as the age or crown cover increased, there was an increased concentration of points in the upper part of the stand.
However, age explained the variation slightly higher in the stand than crown cover (between 18--38\% and 62--64\% of stand height).
Unlike Aber (\protect\hyperlink{ref-aber79}{1979}) who observed
a stable distribution of foliage at the end point of forest succession, the
vertical distribution of LiDAR returns of the 120-year group displayed a small concentration of returns at 33--44\% of stand height and a small decrease below 23\% of stand height.

The most pronounced differences in vertical distributions of LiDAR returns occurred below 80\% of stand height.
The upper sections of the stand often yielded very
small differences in distribution while the lower section yielded large
variations.
The distribution of LiDAR returns in the upper and lower sections of the stand can heavily dependent on the LiDAR survey parameters and species crown shape (Roussel et al. \protect\hyperlink{ref-roussel_etal17}{2017}, \protect\hyperlink{ref-roussel_etal18}{2018}). This might make these sections more variable across different surveys.
Nonetheless, the full model (Equation \eqref{eq:full}) still explained 20\% of the variability from the ground up to 90\% of the stand height.
The difference in variability explained by the species model compared to the full model was greatest at heights above 72\%, and peaked at around 80\% of the stand height (Figure \ref{fig:fig-rsq}).
One possible explanation for this is that the section encompassing 74--90\% of stand height exhibited considerable differences between aspen and all other species.

A common challenge associated with functional data analyses is making functions comparable, a process called registration (Ramsay et al. \protect\hyperlink{ref-ramsay_etal09}{2009}).
We registered the vertical distributions of LiDAR returns by using the highest measured LiDAR return of each stand.
However, this registration makes the distributions potentially more affected by extremely high LiDAR returns and could explain, in part, why the upper section of the stands had a lower R².
For example, in young stands where there might only be a single remaining mature tree, the vertical distribution of LiDAR returns could be compressed.
Using a height quantile as a registration point, such as 95\% height rather than the highest return, could reduce the chances of compression and improve the vertical distributions.
In addition to compression, the vertical distribution of LiDAR returns can be distorted by strong slopes (Liu et al. \protect\hyperlink{ref-liu_etal17}{2017}).
Furthermore, the occlusion of vegetation in lower stand parts underestimates the density of vegetation.
Some of the noise in the vertical distribution can be mitigated by using correction methods such as voxels, or partly accounting for laser incidence angle, footprint size and pulse density (Wilkes et al. \protect\hyperlink{ref-wilkes_etal16}{2016}; Roussel et al. \protect\hyperlink{ref-roussel_etal17}{2017}, \protect\hyperlink{ref-roussel_etal18}{2018}).

There was a decline in R² between 31\% and 46\% of the stand height, where age explained most of the variance of the full model (Figure \ref{fig:fig-rsq}).
While the reasons for the decline are unknown, this section of stand height is associated with an increase in LiDAR returns in the 120-year-old group.
This vertical section of the stands also seems to be a transition for the spruce group and the balsam fir--paper birch group between lower distributions and distributions centered around 50\% of the stand height.
The decrease in R² might be attributed to the intersection of these two distributions.

The nonparametric graphical test of significance that we used is slightly
liberal because is based on the Freedman-Lane algorithm which uses an approximation from permutations (Mrkvička et al. \protect\hyperlink{ref-mrkvicka_etal19}{2019}).
However, this permutation method is regarded in the literature as one of the best methods when there are confounding variables (Anderson and Robinson \protect\hyperlink{ref-anderson_robinson01}{2001}; Winkler et al. \protect\hyperlink{ref-winkler_etal14}{2014}).
The graphical output from our analyses is advantageous, especially for identifying the sections of rejection.
This graphical interpretation allowed us to determine the relative heights at which the differences in species, crown cover, and age occurred.
When interpretability is desired, we think that this method could complement, and in some cases replace other methods of analysis for the vertical distribution of LiDAR returns, such as the principal component analysis or linear discriminant analysis (Fedrigo et al. \protect\hyperlink{ref-fedrigo_etal19}{2019}).

In a review of tree species classification using remote sensing, Fassnacht et al.~(\protect\hyperlink{ref-fassnacht_etal16}{2016}) noted that most studies
pursued the optimization of classification accuracy and provided little
information on the causal understanding of species discrimination.
Using nonparametric graphical tests of significance can help us understand the evolution of forest structure by highlighting the association of variables with the distribution of reflective material.
In our study, we applied this method to a selection of simple stands (that are even-aged with clear species dominance) to visually understand the effect of each of these factors.
Further work is necessary in order to apply this method to multi-species or irregular stands.

The median stand area in our study area was 9\,ha, which is large compared to other area-based
approaches.
This allowed for the aggregation of LiDAR returns to identify specific
patterns.
Stands are by design the most homogeneous unit of the forest landscape, and vertical LiDAR distributions are more reliable for identifying species when there is a sufficient number of aggregated returns grouped together.
To determine the optimal area for observing these patterns will require additional research.

\hypertarget{conclusion}{%
\section{Conclusion}\label{conclusion}}

Light detection and ranging (LiDAR) vertical distributions can
provide an understanding of the interplay between tree species, crown cover,
and age.
The use of functional generalized linear models combined with
graphical tests of significance enabled us to interpret the differences in distributions caused by multiple variables.
Using airborne LiDAR surveys makes it is possible to identify ecosystems that are at multiple evolutionary stages.
Vertical LiDAR distribution patterns variations can be observed and eventually linked to structural and functional dynamics.
Our results show that individual species feature distinctive
vertical distributions of LiDAR returns that concentrate with crown
cover and rise with age.
Balsam fir and paper birch had similar
vertical distributions of LiDAR returns, as did white spruce and black
spruce.
Aspen was the most unique species, yielding a more uniform
distribution of LiDAR returns and a peak in the upper part of the stand.
The balsam fir and paper birch group exhibited a peak centered at
around 50\% of stand height, while the distributions from the white spruce and black spruces groups were skewed to below 30\% of the stand height. Increases in crown cover concentrated the distributions of all species at around 50\% of the stand height and deflated the distribution below 33\%.
The effect of age was more diffuse across the whole stand height.
Age increase was associated to a gradual displacement of the vertical distribution higher in the stands up to the 70 years.
The distribution of the 90-- and 120-year-old groups then plateaued and slowly declined.
These results could improve our understanding of the evolution of the forest structure in changing conditions and could be used for LiDAR stand-level species classification.

\hypertarget{acknowledgments}{%
\section*{Acknowledgments}\label{acknowledgments}}
\addcontentsline{toc}{section}{Acknowledgments}

We thank Benoît St-Onge, Nicole K. S. Barker, Sébastien Renard,
Christian Roy and Josh Nowak for comments on early versions of this
manuscript; Tomáš Mrkvička for helpful discussions on the transformation of
heteroscedastic data; and Anick Patry, Antoine Leboeuf, Marc-Olivier Lemonde,
and Jean-François Bourdon from
Ministère des Forêts, de la Faune et des Parcs for providing data and
support.
We also thank the two anonymous reviewers for their thoughtful feedback.

\hypertarget{declarations}{%
\section*{Declarations}\label{declarations}}
\addcontentsline{toc}{section}{Declarations}

\hypertarget{funding}{%
\subsection*{Funding}\label{funding}}
\addcontentsline{toc}{subsection}{Funding}

This work was supported by the Ministère des Forêts de la Faune et des
Parcs du Québec through the Fonds de Recherche Québécois sur la Nature
et les Technologies and the Academy of Finland (project numbers 295100 and 327211) under the UNITE flagship ecosystem.

\hypertarget{conflicts-of-interestcompeting-interests}{%
\subsection*{Conflicts of interest/Competing interests}\label{conflicts-of-interestcompeting-interests}}
\addcontentsline{toc}{subsection}{Conflicts of interest/Competing interests}

We declare no conflicts of interest.

\hypertarget{availability-of-data-and-material}{%
\subsection*{Availability of data and material}\label{availability-of-data-and-material}}
\addcontentsline{toc}{subsection}{Availability of data and material}

The original data was shared and is available upon request from the
Gouvernement du Québec. The processed data is available in our
repository \url{https://github.com/etiennebr/vertical-lidar-paper}

\hypertarget{code-availability}{%
\subsection*{Code availability}\label{code-availability}}
\addcontentsline{toc}{subsection}{Code availability}

The code used to perform the analysis is available in a git repository at
\url{https://github.com/etiennebr/vertical-lidar-paper}

\hypertarget{author-contribution-statement}{%
\subsubsection*{Author contribution statement}\label{author-contribution-statement}}
\addcontentsline{toc}{subsubsection}{Author contribution statement}

ER is the primary author of the manuscript. ER, NCC and JB contributed
to the design of the study; the analysis was conducted by ER and MM; and the
redaction was conducted by ER and revised by other authors. All authors
read and approved this version of the manuscript.

\section*{References}

\hypertarget{refs}{}
\leavevmode\hypertarget{ref-aber79}{}%
Aber JD (1979) Foliage-Height Profiles and Succession in Northern Hardwood Forests. Ecology 60:18--23. \url{https://doi.org/10.2307/1936462}

\leavevmode\hypertarget{ref-anderson_robinson01}{}%
Anderson MJ, Robinson J (2001) Permutation Tests for Linear Models. Australian \& New Zealand Journal of Statistics 43:75--88. \url{https://doi.org/10.1111/1467-842X.00156}

\leavevmode\hypertarget{ref-axelsson_etal18}{}%
Axelsson A, Lindberg E, Olsson H (2018) Exploring Multispectral ALS Data for Tree Species Classification. Remote Sensing 10:183. \url{https://doi.org/10.3390/rs10020183}

\leavevmode\hypertarget{ref-bassow_bazzaz97}{}%
Bassow SL, Bazzaz FA (1997) Intra- and inter-specific variation in canopy photosynthesis in a mixed deciduous forest. Oecologia 109:507--515. \url{https://doi.org/10.1007/s004420050111}

\leavevmode\hypertarget{ref-beland_etal19}{}%
Beland M, Parker G, Sparrow B et al (2019) On promoting the use of lidar systems in forest ecosystem research. Forest Ecology and Management 450:117484. \url{https://doi.org/10.1016/j.foreco.2019.117484}

\leavevmode\hypertarget{ref-breusch_pagan79}{}%
Breusch TS, Pagan AR (1979) A Simple Test for Heteroscedasticity and Random Coefficient Variation. Econometrica 47:1287--1294. \url{https://doi.org/10.2307/1911963}

\leavevmode\hypertarget{ref-budei_st-onge18}{}%
Budei BC, St-Onge B (2018) Variability of Multispectral Lidar 3D and Intensity Features with Individual Tree Height and Its Influence on Needleleaf Tree Species Identification. Canadian Journal of Remote Sensing 44:263--286. \url{https://doi.org/10.1080/07038992.2018.1478724}

\leavevmode\hypertarget{ref-budei_etal18}{}%
Budei BC, St-Onge B, Hopkinson C, Audet F-A (2018) Identifying the genus or species of individual trees using a three-wavelength airborne lidar system. Remote Sensing of Environment 204:632--647. \url{https://doi.org/10.1016/j.rse.2017.09.037}

\leavevmode\hypertarget{ref-cao_etal14}{}%
Cao L, Coops NC, Hermosilla T et al (2014) Using Small-Footprint Discrete and Full-Waveform Airborne LiDAR Metrics to Estimate Total Biomass and Biomass Components in Subtropical Forests. Remote Sensing 6:7110--7135. \url{https://doi.org/10.3390/rs6087110}

\leavevmode\hypertarget{ref-coops_etal09}{}%
Coops NC, Varhola A, Bater CW et al (2009) Assessing differences in tree and stand structure following beetle infestation using lidar data. Canadian Journal of Remote Sensing 35:497--508. \url{https://doi.org/10.5589/m10-005}

\leavevmode\hypertarget{ref-coops_etal07}{}%
Coops N, Hilker T, Wulder M et al (2007) Estimating canopy structure of Douglas-fir forest stands from discrete-return LiDAR. Trees - Structure and Function 21:295--310. \url{https://doi.org/10.1007/s00468-006-0119-6}

\leavevmode\hypertarget{ref-crespo-peremarch_etal20}{}%
Crespo-Peremarch P, Fournier RA, Nguyen V-T et al (2020) A comparative assessment of the vertical distribution of forest components using full-waveform airborne, discrete airborne and discrete terrestrial laser scanning data. Forest Ecology and Management 473:118268. \url{https://doi.org/10.1016/j.foreco.2020.118268}

\leavevmode\hypertarget{ref-depury_farquhar97}{}%
De Pury DGG, Farquhar GD (1997) Simple scaling of photosynthesis from leaves to canopies without the errors of big-leaf models. Plant, Cell \& Environment 20:537--557. \url{https://doi.org/10.1111/j.1365-3040.1997.00094.x}

\leavevmode\hypertarget{ref-ellsworth_reich93}{}%
Ellsworth DS, Reich PB (1993) Canopy structure and vertical patterns of photosynthesis and related leaf traits in a deciduous forest. Oecologia 96:169--178. \url{https://doi.org/10.1007/BF00317729}

\leavevmode\hypertarget{ref-falkowski_etal09}{}%
Falkowski MJ, Evans JS, Martinuzzi S et al (2009) Characterizing forest succession with lidar data: An evaluation for the Inland Northwest, USA. Remote Sensing of Environment 113:946--956. \url{https://doi.org/10.1016/j.rse.2009.01.003}

\leavevmode\hypertarget{ref-farrar95a}{}%
Farrar JL (1995) Trees in Canada. Fitzhenry \& Whiteside Ltd., Ottawa

\leavevmode\hypertarget{ref-fassnacht_etal16}{}%
Fassnacht FE, Latifi H, Stereńczak K et al (2016) Review of studies on tree species classification from remotely sensed data. Remote Sensing of Environment 186:64--87. \url{https://doi.org/10.1016/j.rse.2016.08.013}

\leavevmode\hypertarget{ref-fedrigo_etal18}{}%
Fedrigo M, Newnham GJ, Coops NC et al (2018) Predicting temperate forest stand types using only structural profiles from discrete return airborne lidar. ISPRS Journal of Photogrammetry and Remote Sensing 136:106--119. \url{https://doi.org/10.1016/j.isprsjprs.2017.11.018}

\leavevmode\hypertarget{ref-fedrigo_etal19}{}%
Fedrigo M, Stewart SB, Roxburgh SH et al (2019) Predictive Ecosystem Mapping of South-Eastern Australian Temperate Forests Using Lidar-Derived Structural Profiles and Species Distribution Models. Remote Sensing 11:93. \url{https://doi.org/10.3390/rs11010093}

\leavevmode\hypertarget{ref-freedman_lane83}{}%
Freedman D, Lane D (1983) A Nonstochastic Interpretation of Reported Significance Levels. Journal of Business \& Economic Statistics 1:292--298. \url{https://doi.org/10.1080/07350015.1983.10509354}

\leavevmode\hypertarget{ref-gonsamo_etal13}{}%
Gonsamo A, D'odorico P, Pellikka P (2013) Measuring fractional forest canopy element cover and openness -- definitions and methodologies revisited. Oikos 122:1283--1291. \url{https://doi.org/https://doi.org/10.1111/j.1600-0706.2013.00369.x}

\leavevmode\hypertarget{ref-harding_etal01}{}%
Harding DJ, Lefsky MA, Parker GG, Blair JB (2001) Laser altimeter canopy height profiles: methods and validation for closed-canopy, broadleaf forests. Remote Sensing of Environment 76:283--297. \url{https://doi.org/10.1016/S0034-4257(00)00210-8}

\leavevmode\hypertarget{ref-heinzel_koch11}{}%
Heinzel J, Koch B (2011) Exploring full-waveform LiDAR parameters for tree species classification. International Journal of Applied Earth Observation and Geoinformation 13:152--160. \url{https://doi.org/10.1016/j.jag.2010.09.010}

\leavevmode\hypertarget{ref-heinzel_koch12}{}%
Heinzel J, Koch B (2012) Investigating multiple data sources for tree species classification in temperate forest and use for single tree delineation. International Journal of Applied Earth Observation and Geoinformation 18:101--110. \url{https://doi.org/10.1016/j.jag.2012.01.025}

\leavevmode\hypertarget{ref-hilker_etal10}{}%
Hilker T, Leeuwen M van, Coops NC et al (2010) Comparing canopy metrics derived from terrestrial and airborne laser scanning in a Douglas-fir dominated forest stand. Trees 24:819--832. \url{https://doi.org/10.1007/s00468-010-0452-7}

\leavevmode\hypertarget{ref-holmgren_persson04}{}%
Holmgren J, Persson Å (2004) Identifying species of individual trees using airborne laser scanner. Remote Sensing of Environment 90:415--423. \url{https://doi.org/10.1016/S0034-4257(03)00140-8}

\leavevmode\hypertarget{ref-hovi_etal16}{}%
Hovi A, Korhonen L, Vauhkonen J, Korpela I (2016) LiDAR waveform features for tree species classification and their sensitivity to tree- and acquisition related parameters. Remote Sensing of Environment 173:224--237. \url{https://doi.org/10.1016/j.rse.2015.08.019}

\leavevmode\hypertarget{ref-isenburg12}{}%
Isenburg M (2012) LAStools - efficient tools for LiDAR processing. Version 120301URL \url{http://lastools.org}

\leavevmode\hypertarget{ref-karna_etal19}{}%
Karna YK, Penman TD, Aponte C, Bennett LT (2019) Assessing Legacy Effects of Wildfires on the Crown Structure of Fire-Tolerant Eucalypt Trees Using Airborne LiDAR Data. Remote Sensing 11:2433. \url{https://doi.org/10.3390/rs11202433}

\leavevmode\hypertarget{ref-karna_etal20}{}%
Karna YK, Penman TD, Aponte C et al (2020) Persistent changes in the horizontal and vertical canopy structure of fire-tolerant forests after severe fire as quantified using multi-temporal airborne lidar data. Forest Ecology and Management 472:118255. \url{https://doi.org/10.1016/j.foreco.2020.118255}

\leavevmode\hypertarget{ref-kim_etal11}{}%
Kim S, Hinckley T, Briggs D (2011) Classifying individual tree genera using stepwise cluster analysis based on height and intensity metrics derived from airborne laser scanner data. Remote Sensing of Environment 115:3329--3342. \url{https://doi.org/10.1016/j.rse.2011.07.016}

\leavevmode\hypertarget{ref-koenig_hofle16}{}%
Koenig K, Höfle B (2016) Full-Waveform Airborne Laser Scanning in Vegetation Studies---A Review of Point Cloud and Waveform Features for Tree Species Classification. Forests 7:198. \url{https://doi.org/10.3390/f7090198}

\leavevmode\hypertarget{ref-korhonen_etal11}{}%
Korhonen L, Korpela I, Heiskanen J, Maltamo M (2011) Airborne discrete-return LIDAR data in the estimation of vertical canopy cover, angular canopy closure and leaf area index. Remote Sensing of Environment 115:1065--1080. \url{https://doi.org/10.1016/j.rse.2010.12.011}

\leavevmode\hypertarget{ref-lacointe00}{}%
Lacointe A (2000) Carbon allocation among tree organs: a review of basic processes and representation in functional-structural tree models. Annals of Forest Science 57:521--533. \url{https://doi.org/10.1051/forest:2000139}

\leavevmode\hypertarget{ref-leboeuf_vaillancourt13}{}%
Leboeuf A, Vaillancourt É (2013a) Guide de photo-interprétation des essences forestières du Québec méridional (partie 2 et 3: espèces feuillues et le rehaussement d'image). Bibliothèque et archives nationales du Québec, Québec

\leavevmode\hypertarget{ref-leboeuf_vaillancourt13a}{}%
Leboeuf A, Vaillancourt É (2013b) Guide de photo-interprétation des essences forestières du Québec méridional (partie 1: espèces résineuses). Bibliothèque et archives nationales du Québec, Québec

\leavevmode\hypertarget{ref-lefsky_etal02}{}%
Lefsky MA, Cohen WB, Harding DJ et al (2002) Lidar remote sensing of above-ground biomass in three biomes. Global Ecology and Biogeography 11:393--399. \url{https://doi.org/10.1046/j.1466-822x.2002.00303.x}

\leavevmode\hypertarget{ref-liu_etal17}{}%
Liu J, Skidmore AK, Heurich M, Wang T (2017) Significant effect of topographic normalization of airborne LiDAR data on the retrieval of plant area index profile in mountainous forests. ISPRS Journal of Photogrammetry and Remote Sensing 132:77--87. \url{https://doi.org/10.1016/j.isprsjprs.2017.08.005}

\leavevmode\hypertarget{ref-macarthur_horn69}{}%
MacArthur RH, Horn HS (1969) Foliage Profile by Vertical Measurements. Ecology 50:802--804

\leavevmode\hypertarget{ref-magnussen_etal99}{}%
Magnussen S, Eggermont P, LaRiccia VN (1999) Recovering Tree Heights from Airborne Laser Scanner Data. Forest Science 45:407--422. \url{https://doi.org/https://doi.org/10.1093/forestscience/45.3.407}

\leavevmode\hypertarget{ref-maltamo_etal05}{}%
Maltamo M, Packalén P, Yu X et al (2005) Identifying and quantifying structural characteristics of heterogeneous boreal forests using laser scanner data. Forest Ecology and Management 216:41--50. \url{https://doi.org/10.1016/j.foreco.2005.05.034}

\leavevmode\hypertarget{ref-martin-ducup_etal16}{}%
Martin-Ducup O, Schneider R, Fournier RA (2016) Response of sugar maple (Acer saccharum, Marsh.) tree crown structure to competition in pure versus mixed stands. Forest Ecology and Management 374:20--32. \url{https://doi.org/10.1016/j.foreco.2016.04.047}

\leavevmode\hypertarget{ref-mehtatalo06}{}%
Mehtätalo L (2006) Eliminating the effect of overlapping crowns from aerial inventory estimates. Can J For Res/Rev can rech for 36:1649--1660. \url{https://doi.org/https://doi.org/10.1139/x06-066}

\leavevmode\hypertarget{ref-mrkvicka_etal20}{}%
Mrkvička T, Myllymäki M, Jílek M, Hahn U (2020) A one-way ANOVA test for functional data with graphical interpretation. Kybernetika 432--458. \url{https://doi.org/10.14736/kyb-2020-3-0432}

\leavevmode\hypertarget{ref-mrkvicka_etal19}{}%
Mrkvička T, Roskovec T, Rost M (2019) A Nonparametric Graphical Tests of Significance in Functional GLM. Methodol Comput Appl Probab. \url{https://doi.org/10.1007/s11009-019-09756-y}

\leavevmode\hypertarget{ref-mrnq07}{}%
MRNQ (2007) Norme de photo-interprétation (version provisoire). Direction des inventaires forestiers, Gouvernement du Québec

\leavevmode\hypertarget{ref-muss_etal11}{}%
Muss JD, Mladenoff DJ, Townsend PA (2011) A pseudo-waveform technique to assess forest structure using discrete lidar data. Remote Sensing of Environment 115:824--835. \url{https://doi.org/10.1016/j.rse.2010.11.008}

\leavevmode\hypertarget{ref-myllymaki_mrkvicka20}{}%
Myllymäki M, Mrkvička T (2020) GET: Global envelopes in R. arXiv:191106583 {[}statME{]}

\leavevmode\hypertarget{ref-myllymaki_etal17}{}%
Myllymäki M, Mrkvička T, Grabarnik P et al (2017) Global envelope tests for spatial processes. Journal of the Royal Statistical Society B 79:381--404. \url{https://doi.org/10.1111/rssb.12172}

\leavevmode\hypertarget{ref-palace_etal15}{}%
Palace MW, Sullivan FB, Ducey MJ et al (2015) Estimating forest structure in a tropical forest using field measurements, a synthetic model and discrete return lidar data. Remote Sensing of Environment 161:1--11. \url{https://doi.org/10.1016/j.rse.2015.01.020}

\leavevmode\hypertarget{ref-papa_etal20}{}%
Papa D de A, Almeida DRA de, Silva CA et al (2020) Evaluating tropical forest classification and field sampling stratification from lidar to reduce effort and enable landscape monitoring. Forest Ecology and Management 457:117634. \url{https://doi.org/10.1016/j.foreco.2019.117634}

\leavevmode\hypertarget{ref-parker_etal04}{}%
Parker GG, Harmon ME, Lefsky MA et al (2004) Three-dimensional structure of an old-growth Pseudotsuga-Tsuga canopy and its implications for radiation balance, microclimate, and gas exchange. Ecosystems 7:440--453

\leavevmode\hypertarget{ref-pretzsch_dieler12}{}%
Pretzsch H, Dieler J (2012) Evidence of variant intra- and interspecific scaling of tree crown structure and relevance for allometric theory. Oecologia 169:637--649. \url{https://doi.org/10.1007/s00442-011-2240-5}

\leavevmode\hypertarget{ref-purves_etal07}{}%
Purves DW, Lichstein JW, Pacala SW (2007) Crown Plasticity and Competition for Canopy Space: A New Spatially Implicit Model Parameterized for 250 North American Tree Species. PLoS ONE 2:e870. \url{https://doi.org/10.1371/journal.pone.0000870}

\leavevmode\hypertarget{ref-racine_etal14}{}%
Racine EB, Coops NC, St-Onge B, Bégin J (2014) Estimating Forest Stand Age from LiDAR-Derived Predictors and Nearest Neighbor Imputation. Forest Science 60:128--136. \url{https://doi.org/http://dx.doi.org/10.5849/forsci.12-088}

\leavevmode\hypertarget{ref-ramsay_etal09}{}%
Ramsay J, Hooker G, Graves S (2009) Functional Data Analysis with R and MATLAB. Springer New York, New York, NY

\leavevmode\hypertarget{ref-ramsay_silverman05}{}%
Ramsay J, Silverman BW (2005) Functional data analysis, 2nd edn. Springer, New York

\leavevmode\hypertarget{ref-raty_etal16}{}%
Räty J, Vauhkonen J, Maltamo M, Tokola T (2016) On the potential to predetermine dominant tree species based on sparse-density airborne laser scanning data for improving subsequent predictions of species-specific timber volumes. For Ecosyst 3:1. \url{https://doi.org/10.1186/s40663-016-0060-0}

\leavevmode\hypertarget{ref-rcoreteam20}{}%
R Core Team (2020) R: A Language and Environment for Statistical Computing version 3.6.3. R Foundation for Statistical Computing, Vienna, Austria

\leavevmode\hypertarget{ref-riggins_etal09}{}%
Riggins JJ, Tullis JA, Stephen FM (2009) Per-segment Aboveground Forest Biomass Estimation Using LIDAR-Derived Height Percentile Statistics. GIScience \& Remote Sensing 46:232--248. \url{https://doi.org/10.2747/1548-1603.46.2.232}

\leavevmode\hypertarget{ref-roussel_etal18}{}%
Roussel J-R, Béland M, Caspersen J, Achim A (2018) A mathematical framework to describe the effect of beam incidence angle on metrics derived from airborne LiDAR: The case of forest canopies approaching turbid medium behaviour. Remote Sensing of Environment 209:824--834. \url{https://doi.org/10.1016/j.rse.2017.12.006}

\leavevmode\hypertarget{ref-roussel_etal17}{}%
Roussel J-R, Caspersen J, Béland M et al (2017) Removing bias from LiDAR-based estimates of canopy height: Accounting for the effects of pulse density and footprint size. Remote Sensing of Environment 198:1--16. \url{https://doi.org/10.1016/j.rse.2017.05.032}

\leavevmode\hypertarget{ref-seavy_etal09}{}%
Seavy NE, Viers JH, Wood JK (2009) Riparian bird response to vegetation structure: a multiscale analysis using LiDAR measurements of canopy height. Ecological Applications 19:1848--1857. \url{https://doi.org/10.1890/08-1124.1}

\leavevmode\hypertarget{ref-stark_etal12}{}%
Stark SC, Leitold V, Wu JL et al (2012) Amazon forest carbon dynamics predicted by profiles of canopy leaf area and light environment. Ecology Letters 15:1406--1414. \url{https://doi.org/10.1111/j.1461-0248.2012.01864.x}

\leavevmode\hypertarget{ref-tackenberg07}{}%
Tackenberg O (2007) A New Method for Non-destructive Measurement of Biomass, Growth Rates, Vertical Biomass Distribution and Dry Matter Content Based on Digital Image Analysis. Ann Bot 99:777--783. \url{https://doi.org/10.1093/aob/mcm009}

\leavevmode\hypertarget{ref-thorpe_etal10}{}%
Thorpe HC, Astrup R, Trowbridge A, Coates KD (2010) Competition and tree crowns: A neighborhood analysis of three boreal tree species. Forest Ecology and Management 259:1586--1596. \url{https://doi.org/10.1016/j.foreco.2010.01.035}

\leavevmode\hypertarget{ref-vaughn_etal12}{}%
Vaughn NR, Moskal LM, Turnblom EC (2012) Tree Species Detection Accuracies Using Discrete Point Lidar and Airborne Waveform Lidar. Remote Sensing 4:377--403. \url{https://doi.org/10.3390/rs4020377}

\leavevmode\hypertarget{ref-vierling_etal10}{}%
Vierling KT, Bässler C, Brandl R et al (2010) Spinning a laser web: predicting spider distributions using LiDAR. Ecological Applications 21:577--588. \url{https://doi.org/10.1890/09-2155.1}

\leavevmode\hypertarget{ref-vierling_etal08}{}%
Vierling KT, Vierling LA, Gould WA et al (2008) Lidar: shedding new light on habitat characterization and modeling. Frontiers in Ecology and the Environment 6:90--98. \url{https://doi.org/10.1890/070001}

\leavevmode\hypertarget{ref-white_etal13}{}%
White JC, Wulder MA, Varhola A et al (2013) A best practices guide for generating forest inventory attributes from airborne laser scanning data using an area-based approach. The Forestry Chronicle 89:722--723. \url{https://doi.org/10.5558/tfc2013-132}

\leavevmode\hypertarget{ref-wilkes_etal16}{}%
Wilkes P, Jones SD, Suarez L et al (2016) Using discrete-return airborne laser scanning to quantify number of canopy strata across diverse forest types. Methods in Ecology and Evolution 7:700--712. \url{https://doi.org/https://doi.org/10.1111/2041-210X.12510}

\leavevmode\hypertarget{ref-winkler_etal14}{}%
Winkler AM, Ridgway GR, Webster MA et al (2014) Permutation inference for the general linear model. NeuroImage 92:381--397. \url{https://doi.org/10.1016/j.neuroimage.2014.01.060}

\leavevmode\hypertarget{ref-wulder_etal12}{}%
Wulder MA, White JC, Nelson RF et al (2012) Lidar sampling for large-area forest characterization: A review. Remote Sensing of Environment 121:196--209. \url{https://doi.org/10.1016/j.rse.2012.02.001}

\leavevmode\hypertarget{ref-orka_etal09}{}%
Ørka HO, Næsset E, Bollandsås OM (2009) Classifying species of individual trees by intensity and structure features derived from airborne laser scanner data. Remote Sensing of Environment 113:1163--1174. \url{https://doi.org/10.1016/j.rse.2009.02.002}

\end{document}